\documentclass[twocolumn]{aastex62}

\usepackage[binary-units]{siunitx}
\usepackage[natbib]{}
\usepackage{graphicx}
\usepackage{bm}
\usepackage{mathtools}
\usepackage{siunitx}
\usepackage{enumitem}

\usepackage[flushleft]{threeparttable}

\graphicspath{{./}{figures/}}

\shorttitle{AMISS II: The CO(2--1)/CO(1--0) Line Ratio}
\shortauthors{Keenan et al.}

\defcitealias{keenan+24a}{Paper I}

\begin{document}

\title{The Arizona Molecular ISM Survey with the SMT:\\
Variations in the CO(2--1)/CO(1--0) Line Ratio Across the Galaxy Population}

\correspondingauthor{R. P. Keenan}
\author[0000-0003-1859-9640]{Ryan P. Keenan}
\affiliation{Steward Observatory, University of Arizona, 933 North Cherry Avenue, Tucson, AZ 85721, USA}
\affiliation{Max-Planck-Institut für Astronomie, Königstuhl 17, D-69117 Heidelberg, Germany}
\email{keenan@mpia.de}

\author[0000-0002-2367-1080]{Daniel P. Marrone}
\affiliation{Steward Observatory, University of Arizona, 933 North Cherry Avenue, Tucson, AZ 85721, USA}

\author[0000-0002-3490-146X]{Garrett K. Keating}
\affiliation{Center for Astrophysics, Harvard \& Smithsonian, 60 Garden Street, Cambridge, MA 02138, USA}

\begin{abstract}

The $J=1\rightarrow0$ spectral line of carbon monoxide (CO(1--0)) is the canonical tracer of molecular gas. However, CO(2--1) is frequently used in its place, following the assumption that the higher energy line can be used to infer the CO(1--0) luminosity and molecular gas mass. The use of CO(2--1) depends on a knowledge of the ratio between CO(2--1) and CO(1--0) luminosities, $r_{21}$. Here we present galaxy-integrated $r_{21}$ measurements for 122 galaxies spanning stellar masses from $10^9$ to $10^{11.5}$~M$_\odot$ and star formation rates (SFRs) from $0.08$ to $35$~M$_\odot$~yr$^{-1}$. We find strong trends between $r_{21}$ and SFR, SFR surface density, star formation efficiency, and distance from the star formation main sequence (SFMS). 
We show that the assumption of a constant $r_{21}$ can introduce biases into the molecular gas trends in galaxy population studies and demonstrate how this affects the recovery of important galaxy scaling relations, including the Kennicutt-Schmidt law and the relation between SFMS offset and star formation efficiency. We provide a prescription which accounts for variations in $r_{21}$ as a function of SFR and can be used to convert between CO(2--1) and CO(1--0) when only one line is available. Our prescription matches variations in $r_{21}$ for both AMISS and literature samples and can be used to derive more accurate gas masses from CO(2--1) observations.
\end{abstract}



\section{Introduction} \label{sec:intro}

Molecular hydrogen (H$_2$) is the primary fuel for the formation of new stars. This results in a tight link between star formation rate (SFR) and the abundance of molecular gas, both locally and on the scale of whole galaxies \citep{kennicutt+12,kennicutt98}. In practice, the abundance of H$_2$ is difficult to measure directly. H$_2$ emission lines are not excited at the low temperatures characteristic of molecular clouds and do not trace the total gas mass, while $H_2$ absorption studies are restricted to sightlines with bright background sources. Instead, emission from rotational transitions of $^{12}$C$^{16}$O carbon monoxide -- the second most abundant interstellar molecule -- is the preferred tracer of molecular gas abundance. 

The fundamental $J=1\rightarrow0$ transition of $^{12}$C$^{16}$O (hereafter CO(1--0)) is easily excited under conditions prevalent in molecular clouds. For typical clouds the luminosity of this line can be converted to a molecular gas mass via a mass to light ratio $\alpha_{\rm CO}$ \citep{bolatto+13}, although the conversion factor may drop in the warmer gas conditions of ultraluminous infrared galaxies \citep[][but see \citealt{dunne+22}]{downes+98} or galaxy centers \citep{sandstrom+13,denbrok+23} and may increase in low metallicity regions \citep{carleton+17,accurso+17}. Studies of CO(1--0) have revealed tight correlations between SFR and molecular gas mass (or the surface density of these quantities) both in resolved regions \citep{leroy+08,bigiel+08,bigiel+11,sanchez+21} and integrated over whole galaxies \citep{kennicutt98,delosreyes+19,kennicutt+21}, characterized by a single scaling relation over five orders of magnitude in star formation rate surface density.

Improving millimeter and sub-millimeter observing capabilities have led to extensive observational use of higher $J$ CO transitions. It is now common practice to measure molecular gas masses by observing one of these higher $J$ lines and converting to a CO(1--0) luminosity by assuming a luminosity ratio between the two lines $r_{J1}$. Detecting CO(2--1) tends to require less integration time than CO(1--0), making it a frequent target of extragalactic surveys \citep{bothwell+14,cairns+19,leroy+21}. In studies at higher redshift, the use of CO(2--1) and higher $J$ lines is particularly common as CO(1--0) cannot be observed from the ground at redshifts $0.7\lesssim z\lesssim 1.6$ and is relatively faint even where it can be observed \citep{daddi+10,bauermeister+13b,carilli+13,tacconi+13,freundlich+19,boogaard+19,valentino+20}.

However, the luminosity ratios between CO(1--0) and the $J>1$ lines depend on density, temperature, and optical depth of the emitting gas. Using a constant $r_{J1}$ relies on the assumption that the range of cloud conditions is intrinsically narrow or that ensembles of clouds in varying states combine to produce constant line ratios when averaging over sufficiently large scales. In practice, variations in the $r_{21}$ ratio have long been found in maps of CO in the Milky Way \citep{sakamoto+97,sawada+01}, and more recently in maps of nearby star forming galaxies \citep{leroy+09,koda+12,vlahakis+13,koda+20,denbrok+21,yajima+21,leroy+22,egusa+22,maeda+23}. These studies find correlations between $r_{21}$ and quantities describing the local intensity of star formation. Galaxy-integrated observations have also found correlations between excitation of the $J\geq3$ CO lines and SFR, specific star formation rate, star formation efficiency, interstellar radiation field, and dust temperature \citep{kamenetzky+16,lamperti+20,boogaard+20,liu+21,leroy+22}, with denser, warmer gas exposed to stronger radiation fields showing higher line ratios.

Variability in CO line ratios has non-trivial effects on our understanding of star formation and galaxy evolution. Studies of the star formation law conducted using CO(2--1) \citep{bigiel+11,leroy+13} and higher transitions \citep{komugi+07,iono+09,liu+15,kamenetzky+16,leroy+23} tend to find more linear correlations than studies conducted using CO(1--0). Variations in $r_{J1}$ with SFR provide a natural explanation for this result \citep{narayanan+11,momose+12,moromkuma-matsui+17}. In maps of nearby disk galaxies, \citet{yajima+21} find that using CO(2--1) results in underestimates of the index for the spatially resolved Kennicutt-Schmidt relation by 10-39\% compared to CO(1--0). 

A more justified approach to using $J>1$ CO lines as global molecular gas mass tracers would be to connect variations in the aggregate CO line excitation (and underlying cloud properties) to measurable variations in galaxy properties. On a galaxy-integrated scale, this can be done by measuring $r_{J1}$ for galaxies spanning wide ranges in global properties such as mass and SFR, and fitting scaling relations to describe the CO line excitation. 

To date such investigations have been limited by small sample sizes, narrow dynamic range in galaxy properties, and complicated or biased target selection. Further, the calibration of millimeter data is difficult, and even high quality datasets from modern observatories can be subject to uncertainties at the 10-20\% level. This can serve as the limiting uncertainty for line ratios measured using multiple facilities or when synthesizing literature data spanning several decades of evolving instrumentation and calibration practices \citep{denbrok+21}. \citet{leroy+22} find that calibration uncertainties are comparable in magnitude to the variations in low-$J$ line ratios, making it particularly difficult to fit scaling relations or extrapolate results to other datasets.

To address these challenges, we have carried out the Arizona Molecular ISM Survey with the SMT (AMISS), a multi-line CO survey focused on providing uniform and well-calibrated line ratio measurements for a large sample of $z\sim0$ galaxies. Here we use the AMISS line catalog to construct the largest uniform sample of galaxy-integrated $r_{21}$ measurements to date. Our galaxy sample covers a wide range in stellar mass ($10^{9.0}<M_*<10^{11.5}$ M$_\odot$) and SFR ($10^{-3}<{\rm SFR}<10^{1.5}$ M$_\odot$~yr$^{-1}$ with CO detections for $10^{-1.1}<{\rm SFR}<10^{1.5}$ M$_\odot$~yr$^{-1}$). All data for each CO line was gathered with a single telescope, and we have carefully characterized the statistical and systematic uncertainties affecting our measurements \citep{keenan+24a}.
For the first time, this sample allows us to directly measure galaxy-integrated correlations between $r_{21}$ and galaxy properties like SFR at high statistical significance. 

The remainder of this paper is structured as follows: in Section~\ref{sec:survey} we describe our sample and pertinent aspects of the observations and data reduction. In Section~\ref{sec:results} we present our measurements of $r_{21}$ and study how $r_{21}$ varies as a function of galaxy properties. In Section~\ref{sec:discussion} we discus the physical meaning of trends between $r_{21}$ and star formation activity, provide a prescription for estimating $r_{21}$ based on measured galaxy properties, and compare our results to previous studies. Section~\ref{sec:var_r21-ks} then explores the consequences of trends in $r_{21}$ for studies addressing the role of molecular gas in galaxy evolution. Our conclusions are briefly summarized in Section~\ref{sec:conclusion}. Throughout this paper we assume a a flat $\Lambda$CDM cosmology with $H_0=70$ and $\Omega_m=0.3$. We use $\log$ to denote base-10 logarithms.


\section{Survey Description}\label{sec:survey}

\begin{figure*}
    \centering
    \includegraphics[width=\textwidth]{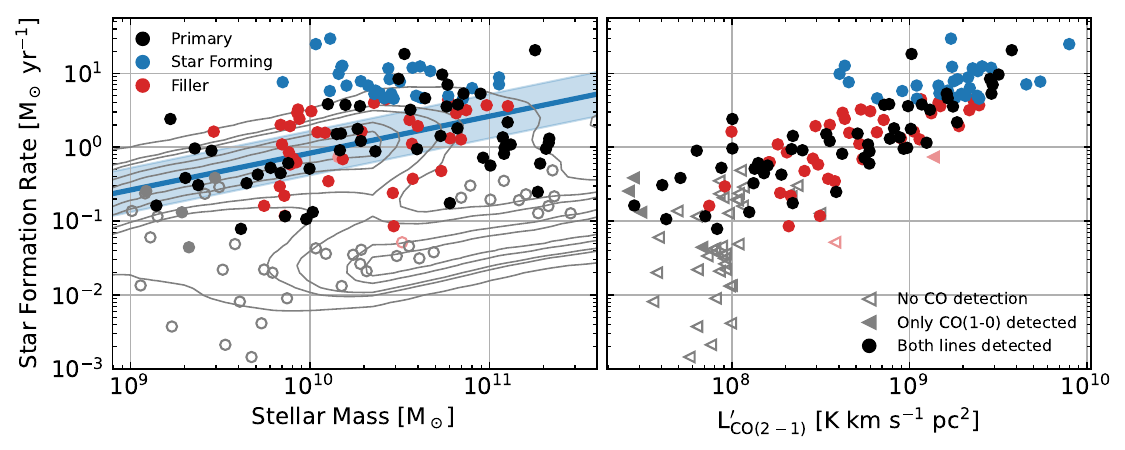}
    \caption{Left: the distribution of the AMISS sample in the stellar mass--SFR plane. Dark colored, filled points show the location of galaxies detected in both CO(2--1) and CO(1--0) and used in this study. Lighter filled points show the locations of galaxies detected in only CO(1--0) and open points show galaxies undetected in both CO lines. Points are colored according to the subsample from which they are drawn -- primary in black, star forming in blue, and filler in red. The blue line and filled region show the star forming main sequence, as parameterized by \citet{speagle+14}, and gray contours show the distribution of SFR at a given mass for all SDSS galaxies at $z<0.05$.
    Right: the distribution of the sample in CO(2--1) luminosity and SFR. Colors and fill styles are the same as in the left panel. Galaxies undetected in CO(2--1) are shown as 3$\sigma$ upper limits (leftward triangles).}
    \label{fig:sample}
\end{figure*}

AMISS was designed to accurately compare CO(2--1) and CO(3--2) to CO(1--0) as tracers of molecular gas in a diverse sample of galaxies. Details of the survey design, execution and data reduction can be found in \citet[][hereafter \citetalias{keenan+24a}]{keenan+24a}; we provide an overview of the most relevant aspects of the survey here.

\subsection{AMISS Observations and CO Data}

Targets for AMISS were drawn from xCOLD GASS \citep{saintonge+17}, which obtained CO(1--0) spectra for 532 SDSS galaxies with a minimum stellar mass of $M_*=10^9$ M$_\odot$ and a redshift range $0.01<z<0.05$. Aside from the mass cutoff, no physical properties of the galaxies were used in selecting targets, making xCOLD GASS a representative sample of the $z\sim0$ galaxy population. AMISS obtained new CO(2--1) observations for three subsamples:
\begin{itemize}[noitemsep,nolistsep]
    \item A ``primary'' sample of 101 galaxies selected to uniformly sample the stellar mass range $10^9<M_*\lesssim10^{11.5}$ M$_\odot$. 
    \item A ``star forming'' sample consisting of all 34 xCOLD GASS galaxies with an SFR above $4.5$~M$_\odot$~yr$^{-1}$ and not already included in the primary sample. 
    \item A ``filler'' sample of 39 galaxies observed as backup targets. These were primarily selected to have bright CO(1--0) emission, but also included handful of low-SFR galaxies with secure CO(1--0) detections.
\end{itemize}

We observed the selected galaxies in CO(2--1) using the Arizona Radio Observatory's Submillimeter Telescope (SMT) on Mt. Graham. 
Targets were typically observed either until a line was detected or a $3\sigma$ upper limit of $L_{\rm CO(2-1)}^\prime<10^8$~K~km~s$^{-1}$ was reached. For a number of galaxies with CO(1--0) detections in xCOLD~GASS but $L_{\rm CO(2-1)}^\prime<10^8$~K~km~s$^{-1}$, we carried out deeper observations until the CO(2--1) line was securely detected.

CO(2--1) fluxes for each galaxy were measured by integrating the spectra over windows set by simultaneous inspection of the AMISS CO(2--1) and xCOLD~GASS CO(1--0)
spectra to include all of the emission from both CO lines. Line fluxes were re-extracted from published xCOLD-GASS CO(1--0) spectra using the same windows. Fluxes were corrected to account for emission falling outside the beam using the procedure described in \citet{saintonge+17}, assuming that the optical and CO distributions are exponential disks with the same size and inclination \citep{leroy+09,bolatto+17}. The corrections are typically small, with median (16th--84th percentile range) corrections of 22\% (10--52\%) for CO(1--0) and 11\% (5--27\%) for CO(2--1). This approach has been validated based on resolved CO maps in a number of works \citep{lisenfeld+11,boselli+14,bothwell+14,leroy+21} and tested for our sample in \citetalias{keenan+24a}.
The xCOLD-GASS CO(1--0) fluxes were further corrected to match the flux scale used by the SMT. $r_{21}$ for each galaxy was then derived using the ratio of the beam corrected luminosities in the two lines.

The median statistical uncertainty in our $r_{21}$ values is $\sigma_{r_{21}}/r_{21}=21\%$. We measured the uncertainty in the calibration of SMT data to be 5\%, and we estimate it to be 10\% for the 30m. Scatter between CO and optical sizes (20\%; \citealt{leroy+09}) implies an additional uncertainty in the aperture corrected luminosities of both lines, which propagates to an uncertainty in $r_{21}$ and is 5\% or less for all sources. We add all of these uncertainties in quadrature to determine the total uncertainty for each of our measurements. The planet models used to set our absolute flux scale are uncertain at the $\sim$5\% level, which implies an additional 0.03~dex systematic uncertainty in the $r_{21}$ values and normalization terms for fits reported in Section~\ref{sec:results}.

\subsection{The $r_{21}$ Sample}\label{ss:survey-sample}

Figure~\ref{fig:sample} shows the distribution of the AMISS sample in stellar mass, SFR, and CO(2--1) luminosity, including both detections and non-detections. Across all of our sub-samples both CO(2--1) and CO(1--0) were detected (${\rm SNR}\ge3$) in 127 galaxies. Excluding two filler targets, we reached a 100\% detection rate for both CO(1--0) and CO(2--1) at SFRs above 0.5 M$_\odot$~yr$^{-1}$. We also detected nearly all galaxies with $L_{\rm CO(2-1)}^\prime > 10^8$ K~km~s$^{-1}$~pc$^2$, with a handful of non-detections in the range $1\times 10^8 \lesssim L_{\rm CO(2-1)}^\prime \lesssim 3\times10^8$ K~km~s$^{-1}$~pc$^2$ corresponding to distant galaxies which we were unable to observe to our target depth prior to the end of the survey. These undetected galaxies are unlikely to be intrinsically different from the detected sources at comparable SFRs. Therefore, our sample should give unbiased constraints on the distribution of and variations in $r_{21}$ in galaxies with SFRs greater than 0.5 M$_\odot$~yr$^{-1}$ and CO luminosities greater than $10^8$ K~km~s$^{-1}$~pc$^2$. 

We also detected both CO lines for a less complete sample of galaxies with SFRs between 0.08 and 0.5 M$_\odot$~yr$^{-1}$, and luminosities between $3\times10^7$ and $1\times10^8$ K~km~s$^{-1}$~pc$^2$. We include these in our analysis to extend our coverage of the parameter space. These sources may provide an incomplete view of trends in $r_{21}$ at low SFR or low CO luminosity, however we have confirmed that their inclusion or exclusion does not alter our fit results over the parameter ranges where our sample is complete (see Appendix~\ref{ap:crosscheck}).

For the following analysis, we use only the galaxies detected in both CO(1--0) and CO(2--1). Among our sample, galaxies detected in only one line generally do not give constraining limits on $r_{21}$, and their inclusion would not significantly alter the conclusions of this paper. We also exclude three sources without valid size measurements and two merging sources for which we cannot determine reliable aperture corrections. This leaves 122 galaxies for which we are able to determine $r_{21}$.

\subsection{Ancillary Data}\label{ss:survey-ancillary}

We draw ancillary information -- including SFRs, metallicities, stellar masses, optical sizes, and infrared luminosities -- from the public xCOLD~GASS catalog \citep{saintonge+17}. The SFRs are described in \citet{janowiecki+17} -- for sources detected in ultraviolet (GALEX) and mid-infrared (WISE), the sum of ${\rm SFR}_{\rm NUV}$ and ${\rm SFR}_{\rm MIR}$ is used (77\% of our sample), otherwise spectral energy distribution (SED)-based SFRs from \citet{wang+11} are used. The SED-based SFRs have been adjusted by 0.17~dex to match the UV+MIR values. Metallicities are derived from the O3N2 line index of \citet{pettini+04} for star forming galaxies and the \citet{kewley+08} mass-metallicity relation (PPO3N2 calibration) for other galaxy types. 

The xCOLD~GASS catalog provides uncertainties for the UV+MIR SFRs, but does not include uncertainties for most other quantities. Where necessary, we assume typical uncertainties of 0.1~dex on stellar masses \citep{salim+16}, 0.2~dex on SED-based SFRs \citep{salim+07}, 0.5~dex for MPA-JHU SFRs \citep{brinchmann+04},  20\% for IR luminosities, 0.025~dex O3N2 metallicities and 0.1~dex for mass-metallicity relation metallicities \citep{hagedorn+24,kewley+08}, and 0.2" for optical radii.


\section{The CO(2--1)/CO(1--0) Luminosity Ratio}\label{sec:results}

\begin{figure*}
    \centering
    \includegraphics[width=\textwidth]{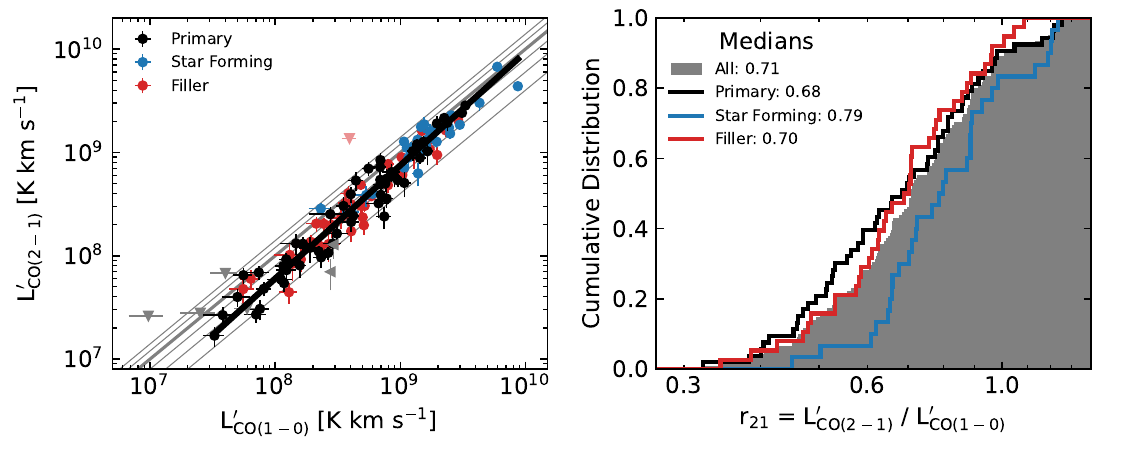}
    \caption{Left: The correlation between CO(1--0) and CO(2--1)luminosities for AMISS galaxies with detections in both CO lines (dark points), along with upper limits in cases where only one line is detected (lighter triangles). Colors indicate which subsample each galaxy belongs to -- primary in black, star forming in blue, and filler in red. Gray diagonal lines corresponds to a line ratios of $r_{21}=0.4$, 0.6, 0.8, 1.0 (thicker), 1.2, 1.4, and 1.6. The black line shows the best power law fit to the data, which is super-linear at more than $4\sigma$ significance. The $1\sigma$ uncertainty region for the fit is comparable to the thickness of the line.
    Right: The cumulative distribution of $r_{21}$ for our entire sample (gray) and the three observational subsamples (same colors as left panel). Medians for each sample are listed.}
    \label{fig:r21}
\end{figure*}

The relationship between CO(2--1) and CO(1--0) line luminosities is plotted in Figure~\ref{fig:r21} (left). We fit a power-law relation between the CO(2--1) and CO(1--0) luminosities of the form: \begin{equation}
    \log \frac{L_{\rm CO(2-1)}^\prime}{10^9\ {\rm K\, km\, s^{-1}\, pc^2}} = m \log \frac{L_{\rm CO(1-0)}^\prime}{10^9\ {\rm K\, km\, s^{-1}\, pc^2}} + b + \mathcal{N}(\sigma),
\end{equation}
where $\mathcal{N}$ describes a log-normal intrinsic scatter of width $\sigma$. We find $m=(1.096\pm0.023)$, $b=(-0.125\pm0.010)$ and $\sigma=(0.05\pm0.02)$. Restricting the fit to only sources with $L_{\rm CO(2-1)}^\prime>10^8$ K~km~s$^{-1}$~pc$^1$ gives consistent results, indicating that detection bias in the luminosity range where our sample is incomplete does not skew our fits. The scaling between $L_{\rm CO(1-0)}^\prime$ and $L_{\rm CO(2-1)}^\prime$ is non-linear at $4.2\sigma$ significance. 

We compute $r_{21}$ for each galaxy in our sample as the ratio of the beam-corrected CO(2--1) and CO(1--0) luminosities:
\begin{equation}
    r_{21} = \frac{L_{\rm CO(2-1)}^\prime}{L_{\rm CO(1-0)}^\prime}.
\end{equation}
With luminosities expressed in units of K~km~s$^{-1}$, $r_{21}$ approaches unity for optically thick gas in local thermodynamic equilibrium. Some studies compute line $r_{21}$ as the ratio of (velocity-integrated) CO fluxes. For comparison to ratios computed with these units (Jy~km~s$^{-1}$), our results should be multiplied by $\nu_{CO(2-1)}^2/\nu_{CO(1-0)}^2 = 4$.

We present ``typical'' values of $r_{21}$ for our sample in Section~\ref{ss:results-r21}, and then explore the dependence of the line ratio on galaxy properties in Section~\ref{ss:results-fit}.

\subsection{The Distribution of $r_{21}$}\label{ss:results-r21}

The right panel of Figure~\ref{fig:r21} shows the distribution of $r_{21}$ values for our entire sample and the three observational subsamples. Considering all 122 galaxies together, we find a median $r_{21}$ of $0.71$ and a 16th to 84th percentile range of $0.51$-$0.95$. 

Within our sub-samples we find a median $r_{21}$ of $0.68$ 
for the primary (mass selected) sample, and $0.79$ 
for the star forming sample. The difference between the samples points to variations in $r_{21}$ across the galaxy population, and indeed we find significant correlation between $r_{21}$ and SFR in the next subsection.

\subsection{Correlations of $r_{21}$ with Galaxy Properties}\label{ss:results-fit}

\begin{figure*}
    \centering
    \includegraphics[width=0.49\textwidth]{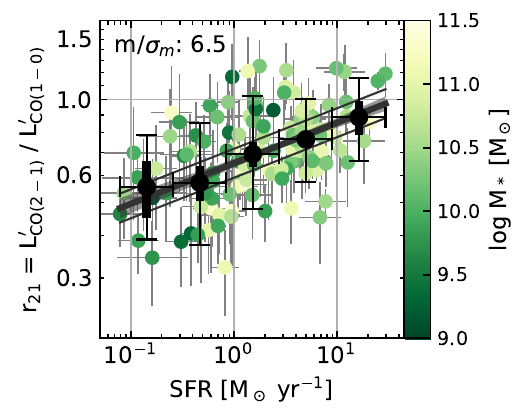}
    \includegraphics[width=0.49\textwidth]{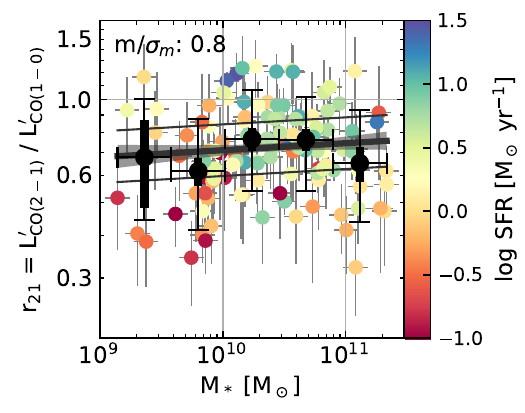}
    \caption{Left: $r_{21}$ as a function SFR. Individual galaxies are colored by stellar mass. Black points show median values in bins of SFR. Thin vertical bars show the 16th-84th percentile range of each bin, and thick vertical bars show the uncertainty in the median. Horizontal bars show the SFR range of each bin. The thick line and filled region show a power law fit to the $r_{21}$--SFR correlation and its $1\sigma$ uncertainty. Thin lines show the best fitting intrinsic scatter in the relationship. 
    Right: Similar to the left panel, but with $r_{21}$ plotted, binned, and fit as a function of stellar mass and galaxies color coded by SFR.}
    \label{fig:r21sfr}
\end{figure*}

Figure~\ref{fig:r21sfr} shows our $r_{21}$ values as a function of SFR and stellar mass. Together, these two plots give a sense of how $r_{21}$ varies. We find a trend of increasing $r_{21}$ with SFR, and no apparent trend between $r_{21}$ and stellar mass. We interpret this to mean that, among the population of $z\sim0$ galaxies studied here, $r_{21}$ is connected more closely to the current star formation activity than structural parameters such as mass.

The uncertainties for individual $r_{21}$ values are significant, resulting in a large apparent scatter in Figure~\ref{fig:r21sfr}. To better illustrate trends, we also show $r_{21}$ for bins along the $x$-axes. To compute these values, we divide our sample into five equally spaced bins, and compute the median $r_{21}$ of all galaxies falling within each bin. We evaluate the uncertainty of these medians by repeatedly re-sampling with replacement from our full set of $r_{21}$ measurements, perturbing each galaxy property and $r_{21}$ value within their respective error distributions, remeasuring the median, and determining the spread of the resulting values.

We also fit power-law scaling relations between $r_{21}$ and variable $x$:
\begin{equation}\label{eq:fit}
    \log r_{21} = m \log x + b + \mathcal{N}(\sigma).
\end{equation}
We find the best fitting parameters and their distribution using a Markov chain Monte Carlo implementation of orthogonal distance regression \citep{hogg+10}. For fits where the $x$-variable is derived from CO luminosity, we determine the covariance between the $x$ and $r_{21}$ errors for each data point by repeatedly drawing new values of $L_{\rm CO(1-0)}^\prime$, $L_{\rm CO(2-1)}^\prime$, and any parameters needed to compute $x$ from their respective error distributions, recomputing $x$ and $r_{21}$, and computing the covariance between realizations of $x$ and $r_{21}$. 

A power law is not the only functional form we could consider, but it is sufficient for our goal of providing a simple empirical means of relating $r_{21}$ to other galaxy properties. Because of the narrow dynamic range spanned by $r_{21}$, the particular choice of functional form has little effect, with alternative parameterizations (e.g., a log-linear relation), giving very similar results. For the properties with which $r_{21}$ shows the strongest correlations, the best fitting power law provides an excellent match to the binned data.\footnote{
Gas physics ultimately sets a limited range of possible line ratios, for instance, $r_{21}$ will saturate at unity for optically thick clouds, implying that any power law fit will eventually extrapolate to nonphysical $r_{21}$ values \citep{leroy+22}. We do not reach this regime with our sample.}

\begin{figure*}
    \centering
    \includegraphics[width=\textwidth]{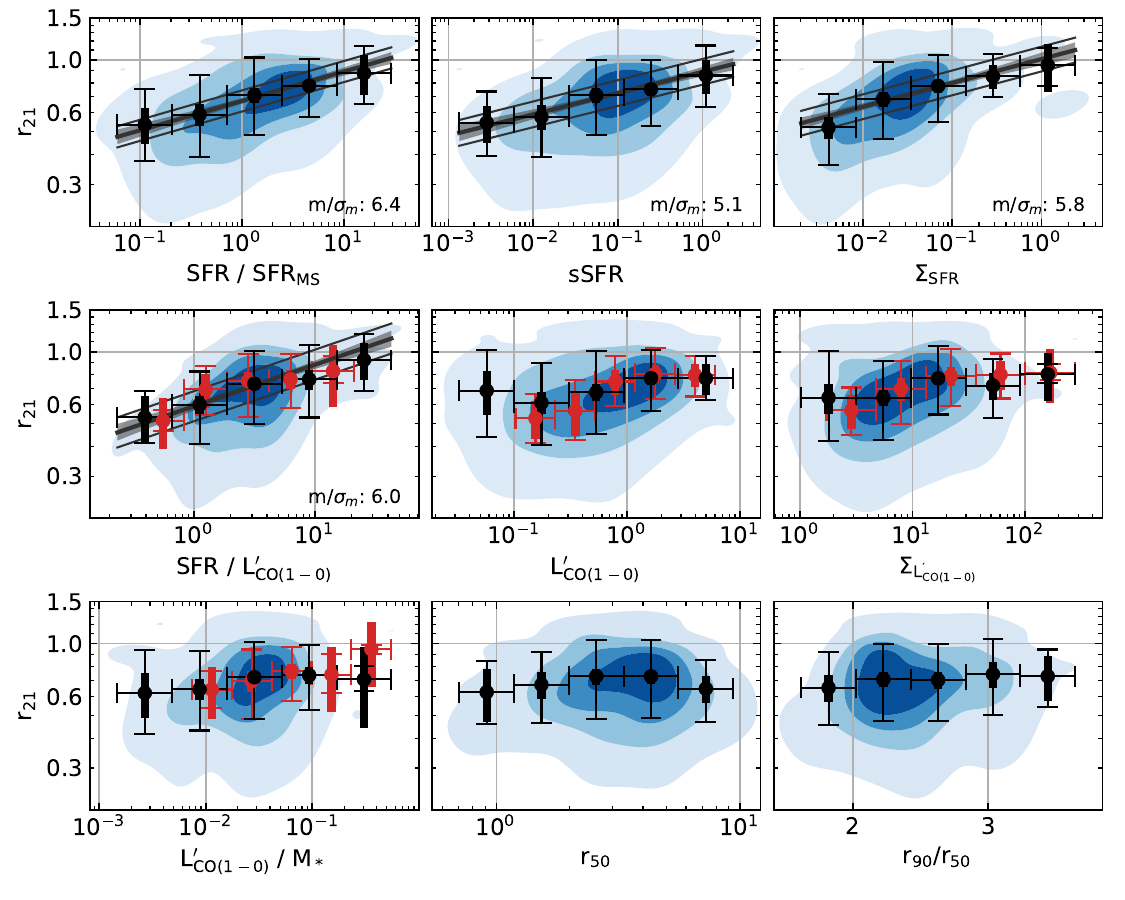}
    \caption{Correlations between $r_{21}$ and a range of galaxy properties. Details of galaxy properties in each subfigure can be found in the text. Contours show the distribution of individual galaxies. Black points show the median $r_{21}$ in bins along the $x$-axis. 
    Thin vertical bars show the 16th-84th percentile range of each bin, and thick vertical bars show the uncertainty in the median. Horizontal bars show the SFR range of each bin. In panels where there is a significant correlation, a thick gray line and filled region show a power law fit and its $1\sigma$ uncertainty. Thin lines show the best fitting intrinsic scatter in the relationship. 
    For the ${\rm SFR} / L_{\rm CO(1-0)}^\prime$, $L_{\rm CO(1-0)}^\prime$, $\Sigma_{\rm CO}$ and $L_{\rm CO(1-0)}^\prime/M_*$ trends, we show binned results for the $x$-quantities derived from independent CO(1--0) measurements from the ARO~12m telescope in red (see Appendix~\ref{ap:12m}).
    }
    \label{fig:r21other}
\end{figure*}

In Figure~\ref{fig:r21other} we show correlations between $r_{21}$ and a wide range of galaxy properties. Corresponding correlation coefficients and power-law fit parameters are listed in Table~\ref{tab:corr}.\footnote{We provide Pearson correlation coefficients, $r$, to allow comparison with other studies. Given our large sample size, coefficients above $\sim0.15$ have a nominal $p<0.01$ for a null hypothesis of no correlation, and coefficients of $\sim0.45$ have $p\sim10^{-7}$. However, our data does not follow a normal distribution in the $x$-variables and, in some cases, has correlated $x$- and $y$-errors, meaning that these values should be taken as indicative only. We recommend the significance by which the slope parameter differs from zero as a better indicator of our confidence in correlations.} 
Across all variables investigated we find a general trend that $r_{21}$ correlates with quantities related to the intensity of star formation and shows little to no correlation with the mass, size, or structural parameters describing the galaxy. We discuss each of these correlations in the remainder of this subsection.

\begin{deluxetable*}{l|c|c|cccc|c}
    \tablecaption{Correlation coefficients and regression parameters for $r_{21}$ and galaxy properties\label{tab:corr}}
    \tablehead{
        \colhead{\textit{x}-variable} & \colhead{N} & \colhead{r} & \colhead{m} & \colhead{b} & \colhead{$\sigma$} & \colhead{$\rho_{mb}$} & \colhead{$m/\sigma_m$} \\
        \colhead{(1)} & \colhead{(2)} & \colhead{(3)} & \colhead{(4)} & \colhead{(5)} & \colhead{(6)} & \colhead{(7)} & \colhead{(8)}
    }
    \startdata
        \hline
        \multicolumn{8}{c}{$r_{21}$ Correlations with SFR Derived Quantities}\\ 
        \hline 
        ${\rm SFR}$ [${\rm M_\odot~yr^{-1}}$] & 120 & 0.487 & $0.119_{-0.018}^{+0.018}$ & $-0.187_{-0.012}^{+0.013}$ & $0.042_{-0.019}^{+0.016}$ & -0.62 & 6.5 \\
        ${\rm sSFR}$ [${\rm Gyr^{-1}}$] & 120 & 0.392 & $0.089_{-0.018}^{+0.018}$ & $-0.050_{-0.020}^{+0.020}$ & $0.055_{-0.016}^{+0.015}$ & 0.86 & 5.1 \\
        ${\rm SFR/SFR_{MS}}$ & 120 & 0.476 & $0.123_{-0.019}^{+0.020}$ & $-0.172_{-0.011}^{+0.011}$ & $0.044_{-0.019}^{+0.016}$ & -0.49 & 6.4 \\
        $\Sigma_{\rm SFR}$ [${\rm M_\odot~yr^{-1}~kpc^{-2}}$] & 120 & 0.470 & $0.100_{-0.017}^{+0.018}$ & $0.003_{-0.026}^{+0.027}$ & $0.053_{-0.015}^{+0.014}$ & 0.92 & 5.8 \\
        ${\rm SFR} / L_{\rm CO(1-0)}^\prime$ [$10^{-9}~{\rm M_\odot~yr^{-1}~(K~km~s^{-1}~pc^{2})^{-1}}$] & 120 & 0.474 & $0.173_{-0.029}^{+0.029}$ & $-0.228_{-0.019}^{+0.018}$ & $0.065_{-0.012}^{+0.012}$ & -0.83 & 6.0 \\
        \hline
        \multicolumn{8}{c}{$r_{21}$ Correlations with $L_{\rm CO}^\prime$ Derived Quantities}\\ 
        \hline 
        $L_{\rm CO(1-0)}^\prime$ [$10^9~{\rm K~km~s^{-1}~pc^2}$] & 121 & 0.208 & $0.067_{-0.024}^{+0.024}$ & $-0.133_{-0.012}^{+0.012}$ & $0.067_{-0.015}^{+0.014}$ & 0.26 & 2.8 \\
        $\Sigma_{L_{CO}^\prime}$ [${\rm K~km~s^{-1}}$] & 121 & 0.213 & $0.075_{-0.025}^{+0.025}$ & $-0.220_{-0.029}^{+0.028}$ & $0.064_{-0.016}^{+0.015}$ & -0.92 & 3.0 \\
        $L_{\rm CO(1-0)}^\prime / M_*$ [${\rm K~km~s^{-1}~pc^{2}~M_\odot^{-1}}$] & 121 & 0.166 & $0.071_{-0.030}^{+0.030}$ & $-0.034_{-0.048}^{+0.047}$ & $0.068_{-0.015}^{+0.015}$ & 0.97 & 2.3 \\
        \hline
        \multicolumn{8}{c}{$r_{21}$ Correlations with Mass, Size, Metallicity, etc.}\\ 
        \hline 
        $M_*$ [$10^{10}~{\rm M_\odot}$] & 120 & 0.066 & $0.021_{-0.026}^{+0.025}$ & $-0.149_{-0.015}^{+0.015}$ & $0.077_{-0.013}^{+0.013}$ & -0.65 & 0.8 \\
        $r_{50}$ [${\rm kpc}$] & 121 & 0.027 & $0.002_{-0.061}^{+0.062}$ & $-0.143_{-0.034}^{+0.033}$ & $0.079_{-0.013}^{+0.013}$ & -0.94 & 0.0 \\
        $r_{90}/r_{50}$ & 121 & 0.112 & $0.054_{-0.031}^{+0.032}$ & $-0.273_{-0.079}^{+0.077}$ & $0.075_{-0.014}^{+0.014}$ & -0.99 & 1.7 \\
        $10^{12}~{\rm O/H}$ & 121 & -0.045 & & & & &
    \enddata
    \tablecomments{Columns are (1) $x$-variable used for correlation; (2) number of galaxies considered - our sample contains 122 galaxies $r_{21}$ measurements, but 1--2 objects are excluded in each correlation because of missing data for the $x$-variable; (3) Pearson correlation coefficient for $x$ and $y$ (see text for caveats); (4-6) fit parameters and uncertainties; (7) correlation of the uncertainties in $m$ and $b$ ($\sigma_{mb}^2/\sigma_m \sigma_b$); (8) the ratio between the fitted slope and its uncertainty. \\
    All fits except $r_{90}/r_{50}$ are of the form $\log r_{21} = m \log x + b + \mathcal{N}(\sigma)$ and are performed accounting for uncertainties in both $x$ and $y$. For $r_{90}/r_{50}$ we fit $\log r_{21} = m x + b + \mathcal{N}(\sigma)$.
    \\ The values of $b$ are subject to an additional $0.03$~dex uncertainty due to uncertainty in the planet models used to set the flux scale for each telescope.}
\end{deluxetable*}

\subsubsection{SFR, sSFR, $\Sigma_{\rm SFR}$, and sSFE}\label{sss:r21-sfr}

We find the strongest correlations between $r_{21}$ and properties related to SFR. In addition to SFR itself, we investigate correlations between $r_{21}$ and 1) the offset between a galaxy's SFR and the expected SFR for a galaxy of the same mass on the star forming main sequence, 2) specific star formation rate (${\rm sSFR} = {\rm SFR}/M_*$),
3) star formation rate surface density ($\Sigma_{\rm SFR}$; computed as half the SFR divided by the area within the effective radius), and 4) ${\rm SFR} / L_{\rm CO(1-0)}^\prime$, a proxy for the star formation efficiency (SFE). 

The correlations between $r_{21}$ and ${\rm SFR}$, ${\rm SFR / SFR_{MS}}$, $\Sigma_{\rm SFR}$, and ${\rm SFR} / L_{\rm CO(1-0)}^\prime$ have similar correlation coefficients, similarly significant slopes, and similar intrinsic scatter parameters. Each of these correlations has a non-zero slope at $\sim6\sigma$ significance. As a check for systematic errors, we re-fit the $r_{21}$-SFR correlation using only subsets of our data and/or alternative SFR and $L_{\rm CO}$ measurements. Details of this test are given in Appendix~\ref{ap:crosscheck} -- in brief, strong correlation between $r_{21}$ and SFR is robustly recovered in all of our tests and we find no evidence that sample selection or unaccounted for systematic errors bias our results.

The correlation between $r_{21}$ and sSFR is slightly weaker than the others, although still statistically significant. This is unsurprising -- as $r_{21}$ is uncorrelated with stellar mass, normalizing SFR by mass will increase the statistical uncertainty of the descriptive variable without adding information about $r_{21}$.

The fitted scatter of 0.04 to 0.06~dex implies that $r_{21}$ typically diverges from our fitted relations by only 10--15\% for a given galaxy. The scatter term in our model should be treated with some care, as it is degenerate with over- or under-estimations of systematic uncertainties. As a check, we re-fitted the $r_{21}$--SFR correlation, setting the systematic uncertainty in our measurements to zero, and found $\sigma=0.07$~dex, which provides an upper limit on the intrinsic width of the relation.

The tight scatter in each of these relations implies that $r_{21}$, measured with sufficient precision, could be a valuable diagnostic of ISM conditions in individual galaxies. This may remain difficult in practice owing to the need for high SNR in both CO lines and the limited dynamic range in $r_{21}$ values compared to typical calibration uncertainties for millimeter observations \citep{denbrok+21,leroy+22}. Still, the large size of our sample makes it possible to understand the behavior of $r_{21}$ statistically. 

The correlation between ${\rm SFR} / L_{\rm CO(1-0)}^\prime$ requires special consideration. The dependence of both $r_{21}$ and ${\rm SFR} / L_{\rm CO(1-0)}^\prime$ on $L_{\rm CO(1-0)}^\prime$ could artificially introduce or enhance the correlations between the two variables. Our fitting procedure attempts to account for error covariance, however this depends on accurate estimates of the uncertainties on all quantities. As a check, we recompute ${\rm SFR} / L_{\rm CO(1-0)}^\prime$ using independently observed CO(1--0) data from the ARO 12m telescope (available for 45 of our sources). These ${\rm SFR} / L_{\rm CO(1-0)}^\prime$ values and the $r_{21}$ values derived with IRAM 30m CO(1--0) data are free of correlated errors. The power law parameters for the new fit agree with the result in Table~\ref{tab:corr} within the uncertainties, and we conclude that correlated errors are not the cause of significant $r_{21}$--${\rm SFR} / L_{\rm CO(1-0)}^\prime$ correlation we measure. We give further details on this test in Appendix~\ref{ap:12m}.

\subsubsection{$L_{\rm CO}^\prime$, $\Sigma_{\rm CO}$, and Gas Fraction}\label{sss:r21-gas}

We also investigate the connection between $r_{21}$ and properties connected to gas mass: 1) $L_{\rm CO(1--0)}^\prime$ -- a proxy for total molecular gas mass, 2) the CO luminosity surface density ($\Sigma_{\rm CO}$; computed as half of $L_{\rm CO(1--0)}^\prime$ divided by the area within the effective radius) and 3) the ratio of CO luminosity and stellar mass ($L_{\rm CO(1-0)}^\prime / M_*$) -- a proxy for the molecular gas mass to stellar mass ratio. We find moderate correlations between $r_{21}$ and these quantities, with correlation coefficients of $\sim 0.2$ and slopes which are non-zero at $2$ to $3\sigma$. 

As our $L_{\rm CO(1-0)}^\prime$ and $r_{21}$ measurements are not independent, anti-correlation between the $x$- and $y$-errors can artificially suppress these correlations. In Appendix~\ref{ap:12m} we re-fit these trends using independent CO(1--0) measurements from the ARO 12m Telescope. Although the sample considered there is smaller and spans a narrower range of galaxy properties, we do find evidence of stronger correlations.

\subsubsection{Stellar Mass and Galaxy Size}\label{sss:r21-mass}

We find no evidence of a correlation between $r_{21}$ and stellar mass (Figure~\ref{fig:r21sfr}), suggesting that $r_{21}$ is relatively insensitive to a galaxy's past evolutionary history and gravitational potential. We also explored correlations between $r_{21}$ and optical half light radius and between $r_{21}$ and concentration (the ratio between the 90\% light radius and half light radius), finding no evidence of correlations for either.

\subsubsection{Metallicity}\label{sss:r21-met}

\begin{figure}
    \centering
    \includegraphics[width=0.48\textwidth]{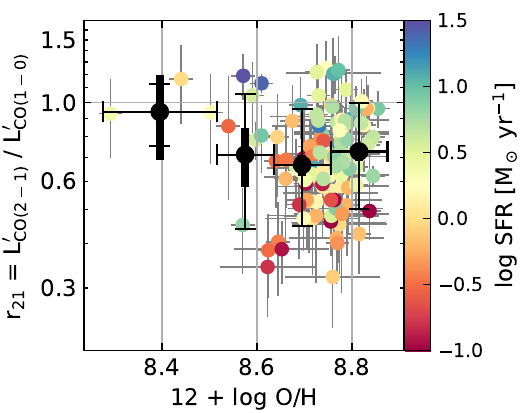}
    \caption{$r_{21}$ as a function metalicity. Individual galaxies are colored by SFR. Black points show median values in bins of metallicity. Thin vertical bars show the 16th-84th percentile range of each bin, and thick vertical bars show the uncertainty in the median. Horizontal bars show the metallicity range of each bin. We have combined the lowest two bins, as the number of objects with $12+\log{\rm O/H}<8.5$ is small.}
    \label{fig:r21met}
\end{figure}

Finally, we explore the relation between $r_{21}$ and metallicity. In low metalicity environments, diffuse regions of molecular clouds may not contain enough dust to shield CO molecules from photodissociation. This means any CO emission that is observed will arise from denser regions of clouds \citep{bolatto+13,accurso+17}, and might imply an anti-correlation between $r_{21}$ and metallicity \citep{penaloza+18}. 

We plot $r_{21}$ versus metallicity in Figure~\ref{fig:r21met}. With one exception, galaxies bellow $12+\log {\rm O/H}\sim8.6$ have $r_{21}\sim1$, seemingly irrespective of their star formation rates. However, the total number of low metallicity systems in our sample is small, and this result should be investigated more fully with targeted samples of low metallicity galaxies.


\section{The Connection Between Molecular Gas Conditions and Star Formation}\label{sec:discussion}

While trends in $r_{21}$ with star formation-related quantities have been reported for resolved regions of galaxies \citep{leroy+09,koda+12,koda+20,denbrok+21,yajima+21,leroy+22,denbrok+23b}, our direct measurement of statistically significant correlations between $r_{21}$ with the properties of star forming galaxies on galaxy-integrated scales is novel. Taken together, the correlations found in Section~\ref{sec:results} between $r_{21}$ and SFR, $\Sigma_{\rm SFR}$, SFE, and sSFR suggest that $r_{21}$ responds to the changes in the star formation activity in a galaxy. 

Other studies have found similar trends for galaxy-integrated line ratios when considering higher energy CO transitions. The CO(3--2) to CO(1--0) ratio, $r_{31}$, has been shown to correlate with SFR, sSFR, and SFE \citep{yao+03,lamperti+20,leroy+22}. \citet{lamperti+20} and \citet{leroy+22} both find the strongest of these correlations to be with SFE (or ${\rm SFR} / L_{\rm CO}^\prime$). \citet{leroy+22} also found tentative evidence of a correlation between galaxy-integrated $r_{21}$ values and ${\rm SFR} / L_{\rm CO}^\prime$, but with low statistical significance. Ratios involving higher energy lines, such as $r_{52}$, also correlate with $\Sigma_{\rm SFR}$ and the strength of the interstellar radiation field, which implies that CO line ratios are determined by the intensity of the radiation falling on the gas clouds \citep{valentino+20,boogaard+20,liu+21}. A resolved study of ratios up to $r_{61}$ in the nearby starburst galaxy NGC 1614 found qualitatively similar correlations between SLED-derived gas temperatures and $\Sigma_{\rm SFR}$ \citep{saito+17}.

Simulations of molecular clouds can help to provide physical meaning to the trends we observe. Molecular clouds are subject to feedback from the stars they form. Hydrodynamic simulations of molecular clouds and cloud complexes in varying environments have shown that stronger interstellar radiation fields and higher cosmic ray reionization rates -- both products of star formation -- lead to higher CO line ratios \citep{penaloza+18,gong+20}. CO molecules at the lower density outskirts of the clouds can be photodissociated in the presence of a strong radiation field, pushing the $\tau=1$ surface for CO radiation deeper into the clouds where higher particle densities result in higher line ratios \citep{penaloza+18}. Clouds in regions -- and perhaps whole galaxies -- with more intense star formation activity will therefore have higher CO line ratios than gas in a more quiescent environment. 

Other environmental factors also play a role. Gas in warm, dense environments such as galaxy centers produces higher line ratios, while low density gas near the transition between atomic and molecular will have a wider range in $r_{21}$ and in particular may have a tail of very low $r_{21}$ values \citep{penaloza+17}. These expectations are generally borne out in resolved observations of star forming disks, which find systematic differences in $r_{21}$ across galactic environments.
On a galaxy-integrated scale, variations in $r_{21}$ may hint at the relative fraction of gas in these different environments. 


\citet{narayanan+14} study CO excitation in lower resolution simulations of galaxy disks and mergers. They find that the molecular gas in starburst galaxies tends to lie at densities and temperatures above the threshold for thermalizing the $J=2\rightarrow1$ transition of CO, while gas in quiescent galaxies lies bellow these thresholds. The physical conditions which set the CO line ratios -- namely density, temperature, and optical depth -- are well correlated with star formation rate surface density in their models. The trend we find between $r_{21}$ and $\Sigma_{\rm SFR}$ is in qualitative agreement with these findings. In detail, we find considerably larger variations in $r_{21}$ with $\Sigma_{\rm SFR}$ than are suggested by their fitting functions which do not extend bellow $\Sigma_{\rm SFR}=0.015$~M$_\odot$~yr$^{-1}$~kpc$^{-2}$ and $r_{21}=0.7$ (see Figure~\ref{fig:r21lit_sigsfr}). Our sample includes a significant number of galaxies lying at lower $\Sigma_{\rm SFR}$ than their models, which likely gives us a better handle on the behavior of $r_{21}$ in environments with little star formation.

\citet{narayanan+14} suggest that global star formation rate, which might be distributed over a whole galaxy disk or concentrated in a nuclear starburst, should not correlate as strongly CO line ratios as $\Sigma_{\rm SFR}$. They reach this conclusion primarily on the basis of higher $J$ CO lines. On the other hand, we find strong correlations between $r_{21}$ and both SFR and $\Sigma_{\rm SFR}$, with SFR showing a slightly more significant slope, higher correlation coefficient, and smaller intrinsic scatter. A possible explanation is that while higher-$J$ CO emission must arise from very dense, warm environments, CO(1--0) and, to a lesser extent, CO(2--1) trace molecular gas in a wider range of environments, making the ratio of these particular lines more sensitive to non-localized conditions such as the overall reaction of gas in dense clouds capable of forming stars.

\subsection{A Scaling Relation for $r_{21}$}\label{ss:discussion-prescription}

Typical practice for deriving molecular gas masses from CO(2--1) is to convert CO(2--1) to CO(1--0) using an assumed value of $r_{21}$. The gas mass can then be estimated using a CO(1--0) luminosity to molecular gas mass conversion factor, $\alpha_{\rm CO(1-0)}$ \citep{bolatto+13}. While a constant $r_{21}$ is often chosen, the results of Section~\ref{sec:results} show that there are appreciable, systematic variations in $r_{21}$ across the galaxy population. A more sophisticated prescription for estimating $r_{21}$ could therefore deliver better molecular gas mass estimates.

The $r_{21}$ ratio is tightly correlated with SFR, $\Sigma_{\rm SFR}$, ${\rm SFR} / L_{\rm CO(1--0)}^\prime$, and ${\rm SFR/SFR_{MS}}$, all to a similar degree. Depending on which ancillary data are best measured, any of these parameters can be used to infer $r_{21}$ using the fits from Table~\ref{tab:corr}. Because SFR is likely to be the simplest to obtain and most precisely measured, here we present a prescription based on the $r_{21}$--SFR correlation. The scaling relations determined in Section~\ref{ss:results-fit} are valid for SFRs from $\sim0.1$ to 35~M$_\odot$~yr$^{-1}$. Above this range, $r_{21}$ is expected to saturate at 1 for optically thick gas \citep{leroy+22}. Observations of ultraluminous infrared galaxies find typical $r_{21}$ of $\sim 1$ \citep{papadopoulos+12, montoya-arroyave+23}, consistent with this expectation. We therefore recommend
\begin{equation}\label{eq:r21prescription}
    \log r_{21} =     
    \begin{dcases}
        0.12 \log {\rm SFR} - 0.19 & -1 \lesssim \log {\rm SFR} < 1.58 \\
        0.0 & 1.58 < \log {\rm SFR}
    \end{dcases}\,.
\end{equation}
The lower panels of Figure~\ref{fig:r21lit} show this prescription compared to the AMISS data and $r_{21}$ measurements compiled from previous literature (discussed in the next section). The intrinsic variations in $r_{21}$ at a given SFR are larger than the uncertainty in the fit over the full range of our data. We therefore recommend adopting a statistical uncertainty of 0.04~dex for $r_{21}$ values derived using Equation~\ref{eq:r21prescription}. We cannot constrain $r_{21}$ below ${\rm SFR}\sim0.1$~M$_\odot$~yr$^{-1}$ based on the AMISS sample, and recommend using caution when extrapolating below this limit.

\subsection{Comparison to Prior Studies: Global $r_{21}$ Values}

\begin{figure*}
    \centering
    \includegraphics[width=\textwidth]{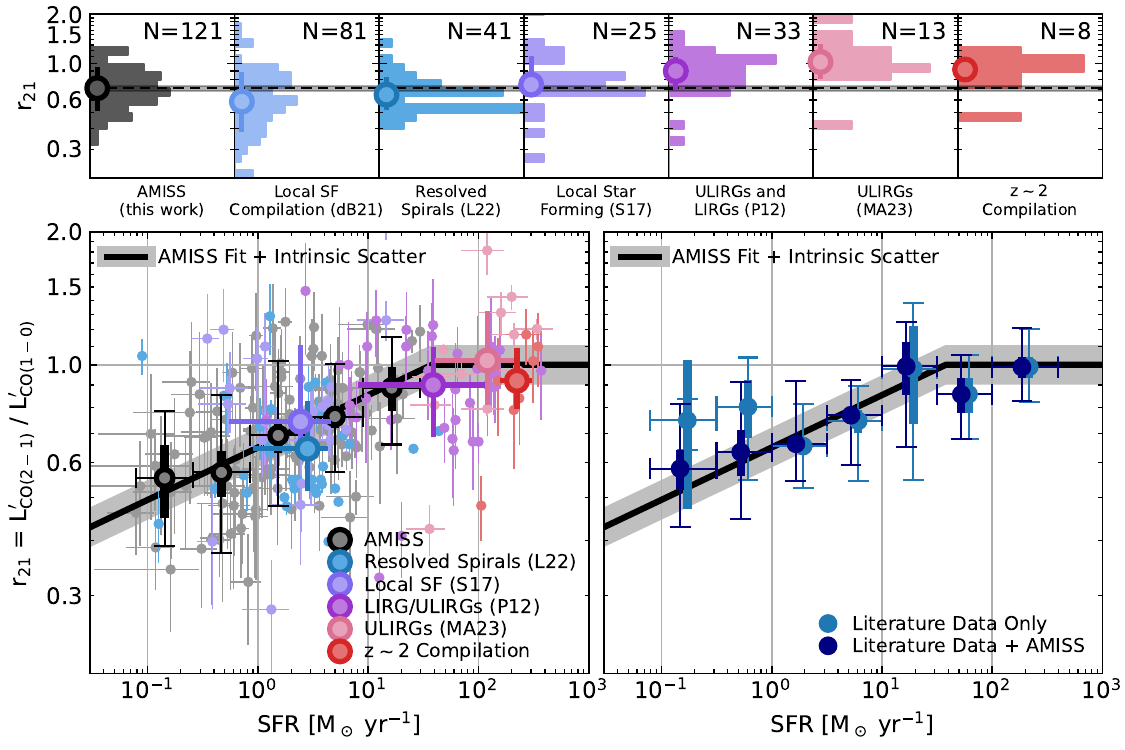}
    \caption{Top: the distribution of $r_{21}$ measurements from AMISS, the compilation by \citet[][disk sample only]{denbrok+21}, resolved nearby galaxies from \citet{leroy+22}, star forming galaxies from \citet{saintonge+17}, LIRGs and ULIRGS from \citet{papadopoulos+12}, ULIRGs from \citet{montoya-arroyave+23}, and a compilation of $1<z<3$ galaxies. Points and error bars show the median and 16th-84th percentile range for each sample. For ease of comparison, the black dashed line and surrounding gray region in all panels show the AMISS median and its $1\sigma$ uncertainty. The number of galaxies in each sample is given in the upper right. Bottom: our prescription for $r_{21}$ as a function of SFR (Equation~\ref{eq:r21prescription}) and its intrinsic scatter are shown by the black line and gray filled region. In the left panel, we reproduce the AMISS galaxies and bins from Figure~\ref{fig:r21sfr} (gray points), and add galaxies from our compilation of literature results (small colored points). Larger colored markers show the median and 16th-84th percentile range for individual samples from the literature. In the right panel we show the compiled data binned by SFR both with (dark blue) and without (lighter blue) the AMISS galaxies. Thin vertical bars show the 16th-84th percentile range of each bin, and thick vertical bars show the uncertainty in the median. Horizontal bars show the SFR range of each bin. While the literature data follows our prescription, the inclusion of the AMISS galaxies is necessary to clearly capture the decrease in $r_{21}$ at low SFRs.}
    \label{fig:r21lit}
\end{figure*}

Prior studies have reported typical values of galaxy-integrated $r_{21}$ ranging from $\sim0.6$ to $\sim1.0$. In Figure~\ref{fig:r21lit} we compare the distribution of $r_{21}$ for the AMISS sample with nearby spiral galaxies from \citet{leroy+22}, star forming galaxies from \citet{saintonge+17}\footnote{We have rescaled the \citet{saintonge+17} data to match our re-calibration of the xCOLD~GASS CO(1--0) fluxes.} and the compilation of \citet{denbrok+21}, IR selected galaxies from \citet{papadopoulos+12} and \citet{montoya-arroyave+23}, and star forming galaxies at cosmic noon \citep[$1<z<3$;][]{daddi+15,bolatto+15,riechers+20}\footnote{\citet{riechers+20} and \citet{bolatto+15} report only $r_{31}$ values for their high redshift galaxies. All of these values are near unity, and we have used them to estimate $r_{21}\simeq r_{31}$ under the assumption that $r_{31}\leq r_{21} \leq 1$}. 

The scatter, both within and between different samples is significant. Trends between SFR and $r_{21}$ are not readily apparent in individual literature samples, however the samples with higher typical SFRs (the IR-selected and cosmic noon galaxies) do tend to show higher $r_{21}$, in qualitative agreement with the tends found in Section~\ref{sss:r21-sfr}. 

AMISS provides a better sampling of the galaxy parameter space than previous studies. Our sample is equal in size to the entire literature compilation of single dish $r_{21}$ measurements presented in \citet{denbrok+21}, but has the advantage of being observed and processed using a uniform methodology. 

We show $r_{21}$ as a function of SFR for the literature sources and AMISS in the lower left panel of Figure~\ref{fig:r21lit}. The medians values of each literature survey lie near the AMISS trend. Much of the variation between literature results can now be clearly attributed to differing galaxy properties of each sample, though differences in facilities, calibration, and methodology across the heterogeneous literature samples likely still play a role. We further explore this by combining all of the literature sources, and computing the median values in bins of SFR. The right panel of Figure~\ref{fig:r21lit} shows the results both with (dark blue) and without (light blue) the AMISS data. At ${\rm SFR}>1~{\rm M_\odot~yr^{-1}}$ the binned literature data closely follows the prescription from Section~\ref{ss:discussion-prescription}. At lower SFRs, the number of literature measurements is small and the median values are uncertain. However, the inclusion of the AMISS data robustly shows the continuation of the decreasing trend in $r_{21}$ at low SFRs.

\subsection{Comparison to Prior Studies: Resolved $r_{21}$ Trends}

\begin{figure*}
    \centering
    \includegraphics[width=\textwidth]{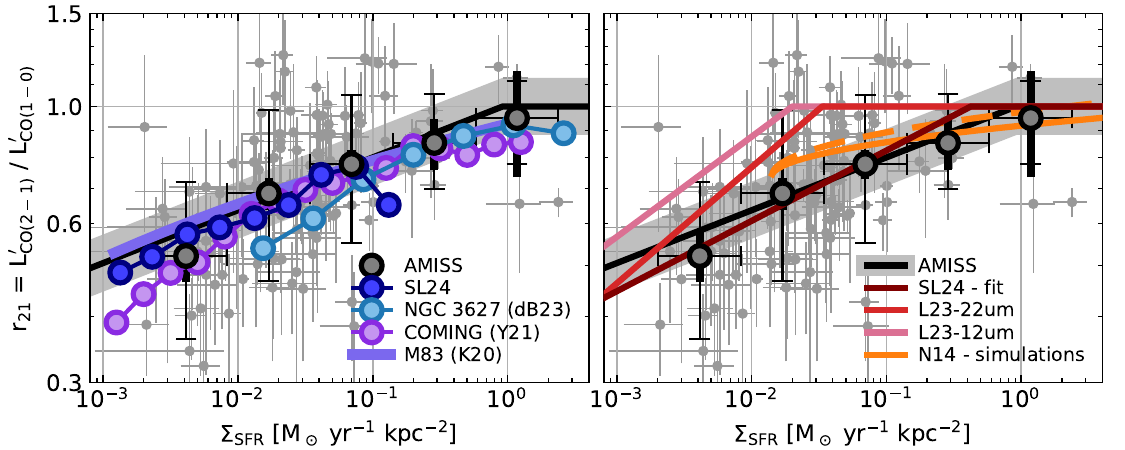}
    \caption{The correlation between $\Sigma_{\rm SFR}$ and $r_{21}$. Light gray points show individual AMISS galaxies, gray points with black circles show the binned AMISS data (reproduced from Figure~\ref{fig:r21other}), and the black line and gray filled region show our fitted power law (truncated at $r_{21}=1$) and $1\sigma$ intrinsic scatter. In the left panel, we compare the AMISS results with resolved $r_{21}$ in bins of $\Sigma_{\rm SFR}$ across individual galaxies or collections of galaxies from \citet{schinnerer+24}, \citet{denbrok+23b}, \citet{yajima+21} and \citet{koda+20}. In the right panel we compare with literature prescriptions for $r_{21}$ from simulations \citep[][orange, resolved and galaxy-integrated pescriptions are shown as solid and dashed lines respectively]{narayanan+14}, ratios of CO--IR correlations \citep[][light red for prescription derived from CO--12$\mu$m correlations, red for prescription derived from CO--22$\mu$m correlations]{leroy+23}, and the fit to a the compilation of resolved $r_{21}$ measurements from \citet[][dark red]{schinnerer+24}.}
    \label{fig:r21lit_sigsfr}
\end{figure*}

The galaxy-integrated correlations between $r_{21}$ and star formation tracers in Table~\ref{tab:corr} are consistent with trends found in resolved regions of nearby galaxies. In a study of nine nearby spirals, \citet{denbrok+21} found slopes ranging from $m=0.08$ to $0.20$ for the correlation between $r_{21}$ and total infrared luminosity surface density $\Sigma_{\rm TIR}$ (a proxy for $\Sigma_{\rm SFR}$), and slopes of $m=0.11$ to $0.24$ between $r_{21}$ and $\Sigma_{\rm TIR}/\Sigma_{\rm CO(1-0)}$ (a proxy for SFE), very similar to our findings for $\Sigma_{\rm SFR}$ and SFR/$L_{\rm CO(1-0)}^\prime$. Similar results are found for other individual galaxies \citep{koda+12,koda+20,denbrok+23b}, and when combining observations of many galaxies \citep{yajima+21,schinnerer+24}. 
In the left panel of Figure~\ref{fig:r21lit_sigsfr} we show the correlation between $r_{21}$ and $\Sigma_{\rm SFR}$ from a number of these studies. The resolved and galaxy-integrated trends are well-matched, with the recent compilation of measurements from \citet{schinnerer+24} lying almost perfectly along the best fit for this relation that we derive in Section~\ref{ss:results-fit}. 

In the right panel of Figure~\ref{fig:r21lit_sigsfr} we show various prescriptions for the $r_{21}$--$\Sigma_{\rm SFR}$ relation from the literature. There is a good match between our result and the recommendation of \citet{schinnerer+24}. On the other hand, prescriptions not directly derived from $r_{21}$ measurements fail to match the observed trends. As discussed in above, the simulation-based predictions of \citet{narayanan+14} give too shallow of a dependence of $r_{21}$ on $\Sigma_{\rm SFR}$. The predictions from \citet{leroy+23} are steeper than our findings or those of the resolved studies. This work used ratios between CO(2--1)-to-infrared and CO(1--0)-to-infrared surface density scaling relations to derive slopes between $r_{21}$ and $\Sigma_{\rm SFR}$ between $0.19$ to $0.26$ (depending on choice of mid-infrared SFR tracer), but did not have access to matched samples of CO(1--0) and CO(2--1) for deriving their predictions. Differences in selection between their CO(1--0) and CO(2--1) samples may explain the mismatch with direct measurements.

Resolved studies find that line ratios vary as a function of galactic environment \citep{koda+12,saito+17,leroy+22,denbrok+23b}. As galaxy-integrated measurements are luminosity- rather than area-weighted, they capture the bulk properties of the molecular gas, but are less sensitive to gas-poor environments that contribute minimally to the total CO luminosity. The general agreement between global and local trends is therefore not trivial. Even among resolved measurements, offsets are observed in the normalization of the $r_{21}$--$\Sigma_{\rm SFR}$ relation when considering different regions \citep{denbrok+23b}. Resolved $r_{21}$ measurements for larger galaxy samples are possible with facilities such as ALMA or NOEMA and will further clarify the connection between global and local CO excitation.

\subsection{Applicability Beyond $z\sim0$}\label{ss:discussion-highz}

\begin{figure*}
    \centering
    \includegraphics[width=\textwidth]{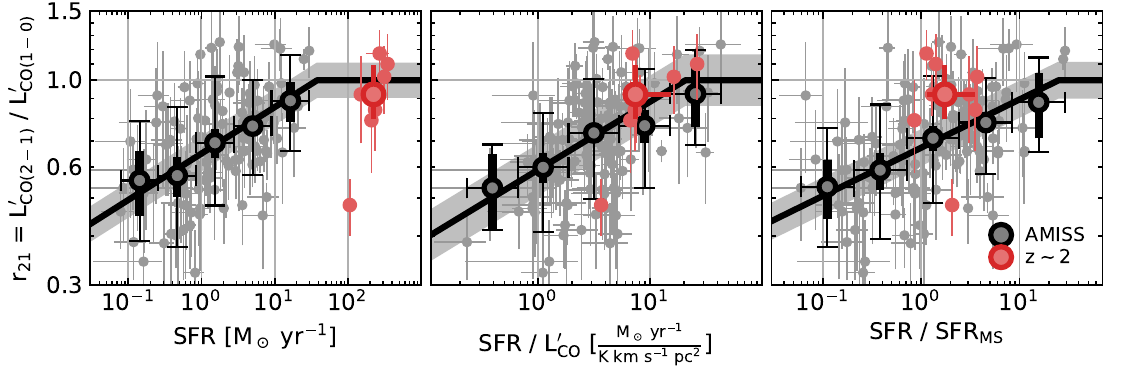}
    \caption{Correlations between $r_{21}$ and SFR (left), ${\rm SFR}/L_{\rm CO(1-0)}^\prime$ (center), and main sequence offset (right) for the AMISS sample (gray) and our compilation of $1<z<3$ galaxies. Lighter points show individual galaxies, while points with dark outlines show bins. For AMISS we reproduce the bins from Figures~\ref{fig:r21} and \ref{fig:r21other}; for the $z\sim2$ galaxies we show a single bin representing the median $x$- and $r_{21}$-values for the sample, with errorbars giving the 16th--84th percentile range. We also show the fitted scaling relations from Table~\ref{tab:corr}, extended to higher values along the $x$-axis assuming that $r_{21}$ saturates at unity.}
    \label{fig:highz}
\end{figure*}

The value of $r_{21}$ in $z\ga 1$ galaxies is of particular interest, as CO(2--1) is often the lowest energy CO transition available for these objects. The samples of such galaxies for which low-$J$ CO line ratios have been measured is growing \citep{daddi+15,bolatto+15,riechers+20,boogaard+20,liu+21,sulzenauer+21}, but remains small with an uneven sampling of redshift, galaxy properties, and CO transitions. Conducting an exhaustive study of $r_{21}$ in distant galaxies remains difficult with current facilities, so it is valuable to consider the insights provided by our $z\sim0$ results.

The SFRs and gas masses of $z\sim2$ main sequence galaxies of a given stellar mass were both approximately an order of magnitude larger at than today \citep{forster_scheiber+20}. This evolution means that not all of the correlations identified in Table~\ref{tab:corr} can extend to galaxies at cosmic noon and beyond. Higher star formation rates combined with the $z\sim0$ correlations of $r_{21}$ with SFR and $\Sigma_{\rm SFR}$, would imply that main sequence galaxies at $z\sim2$ should have $r_{21}\sim1.0$. On the other hand, the $z\sim0$ main sequence offset and star formation efficiency correlations would imply a wider range of $r_{21}$ in $z\sim2$ galaxies, comparable to that seen in the local universe. 

In Figure~\ref{fig:highz} we show $r_{21}$ as a function of SFR, ${\rm SFR} / L_{\rm CO}^\prime$, and main sequence offset for the AMISS sample and our compilation of $z\sim2$ galaxies. The $z>0$ data agree reasonably well with the SFR and ${\rm SFR} / L_{\rm CO}^\prime$ trends (left and center panels), but are systematically separated from the trend with main sequence offset (right panel).
This is consistent with \citet{boogaard+20} who find that the ISM conditions of main-sequence galaxies evolve significantly with redshift.

The small size and limited dynamic range of the $z\sim2$ sample prevents definitive conclusions about how to extend our relations to higher redshift. 
Measuring $r_{21}$ for even a small number low-SFR and/or low-star formation efficiency galaxies at cosmic noon could be valuable for determining which correlations best capture the underlying physical mechanism driving variations in $r_{21}$.


\section{CO(2--1) as Gas Mass Tracer}\label{sec:var_r21-ks}

\begin{figure}
    \centering
    \includegraphics[width=.45\textwidth]{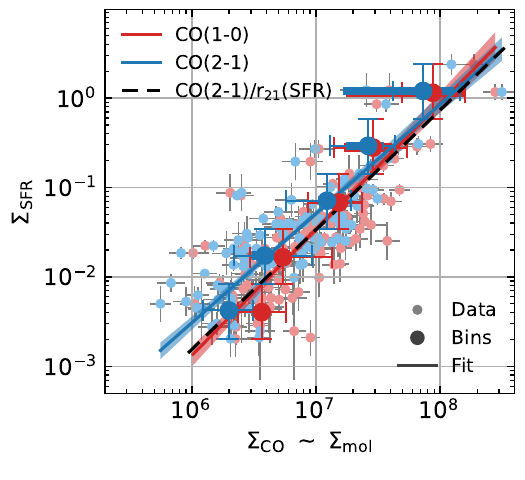}
    \caption{The correlation between CO luminosity surface density and star formation rate surface density for 120 galaxies with both CO(1--0) and CO(2--1) detections from xCOLD~GASS and AMISS. Lighter red and blue points show the distribution of individual galaxies when $\Sigma_{\rm CO}$ is computed using CO(1--0) and CO(2--1) respectively. Darker points show median values for bins along the \textit{$y$-axis}. This binning scheme is chosen so that bins represent the median $\Sigma_{\rm CO(1-0)}$ and $\Sigma_{\rm CO(2-1)}$ of identical sets of galaxies. Thin vertical bars show the limits of each bin, thin horizontal bars show the 16th-84th percentile range of each bin. 
    Lines and filled bands show power law fits. The black dashed line shows a fit to $\Sigma_{\rm CO(2-1)}/r_{21}({\rm SFR})$. The slope for each fit is given in Table~\ref{tab:galaxylaws}.}
    \label{fig:ks}
\end{figure}

\begin{figure*}
    \centering
    \includegraphics[width=\textwidth]{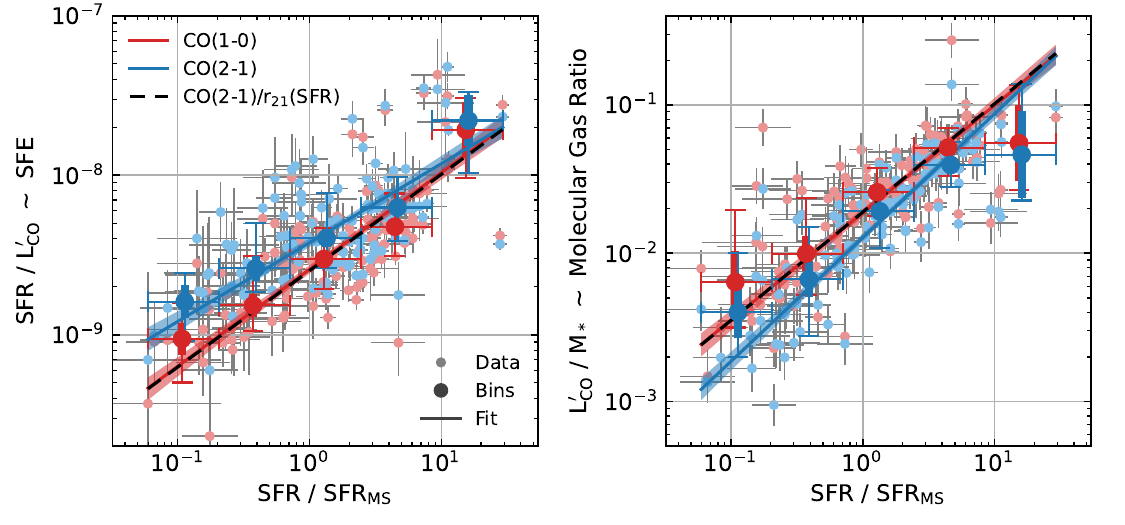}
    \caption{Correlations between ${\rm SFR}/L_{\rm CO}^\prime$ (left) and $L_{\rm CO}^\prime/M_*$ (rigth) with main sequence offset for matched samples galaxies observed in both CO(1--0) and CO(2--1). Lighter points show individual galaxies when CO(1--0) (red) and CO(2--1) (blue) are used in computing the quantity on the $y$-axis. Dark points show median values in bins along the $x$-axis. Each CO(1--0) bin shares an identical set of galaxies with the corresponding CO(2--1) bin. Thin horizontal bars show the extent of each bin in ${\rm SFR}/{\rm SFR_{MS}}$, while thin vertical bars show the 16th-84th percentile range of each bin. Thick vertical bars show the uncertainty of the medians. Lines and filled bands show power law fits and their uncertainties. The black dashed line shows a fit using $L_{\rm CO(2-1)}^\prime/r_{21}({\rm SFR})$. The slope for each fit is given in Table~\ref{tab:galaxylaws}.}
    \label{fig:ms}
\end{figure*}

\begin{deluxetable*}{l|ccc}
    \tablecaption{Power law slopes for scaling relations fit with CO(1-0), CO(2-1), or CO(2-1)/$r_{21}({\rm SFR})$\label{tab:galaxylaws}}
    \tablehead{
        \colhead{} & \multicolumn{3}{c}{Scaling Relation Slope} \\
        \colhead{Scaling Relation} & \colhead{CO(1--0)} & \colhead{CO(2--1)} & \colhead{CO(2--1)/r$_{21}({\rm SFR})$} 
    }
    \startdata
        \hline
        $\log \Sigma_{\rm SFR}$--$\log \Sigma_{\rm CO}$ & $1.41\pm0.10$ & $1.22\pm0.08$ & $1.35\pm0.09$ \\
        $\log {\rm SFR}/{\rm SFR_{MS}}$--$\log {\rm SFR} / L_{\rm CO}$ & $0.61\pm0.05$ & $0.49\pm0.05$ & $0.60\pm0.05$ \\
        $\log {\rm SFR}/{\rm SFR_{MS}}$--$L_{\rm CO} / M_*$ & $0.73\pm0.05$ & $0.83\pm0.05$ & $0.73\pm0.05$
    \enddata
\end{deluxetable*}

The standard practice of deriving gas masses from CO(2--1) by applying a constant value of $r_{21}$ (we will denote a standardized, constant $r_{21}$ as $r_{21}^c$ in the following) and using the CO(1--0) luminosity-to-gas mass conversion factor ($\alpha_{\rm CO}$) implicitly assumes that CO(2--1) and CO(1--0) are tracing the same gas in the same manner. The dynamic range of $r_{21}$ values is small -- our best fit relation changes by only a factor of 2 between SFRs of 0.1 and 30 M$_\odot$~yr$^{-1}$. So for gas mass measurements of an \textit{individual} galaxy this assumption contributes minimally to the overall uncertainty in derived molecular gas mass relative to uncertainty in $\alpha_{\rm CO}$ and, sometimes, the statistical uncertainty in CO luminosity. Using any of the standard $r_{21}^c$ values found in the literature will therefore give a reasonable molecular gas mass estimate. 

However, when studying \textit{variations} in gas properties across samples of galaxies, use of a constant $r_{21}^c$ will bias results in the sense that gas masses of galaxies with little star formation and/or gas will be under-estimated relative to galaxies with more active star formation and more gas. This will alter the slope, and in some cases interpretation, of scaling relations describing the gas properties of galaxies. 

The construction of our sample allows us to directly test how the choice of $r_{21}$ affects conclusions about the scaling relationships between gas mass and galaxy properties. Here, we use a matched sample of galaxies with both CO(1--0) and CO(2--1) luminosity measurements, to investigate how different molecular gas mass estimators affect the index of the Kennicutt-Schmidt law \citep[e.g.][]{kennicutt98,delosreyes+19,kennicutt+21} and the slopes of scaling laws relating gas consumption to location on the star forming main sequence \citep[e.g.][]{saintonge+17,tacconi+18,saintonge+22}.

In Figure~\ref{fig:ks} we show the correlation between $\Sigma_{\rm SFR}$ and $\Sigma_{\rm CO}$, an analogue for the Kennicutt-Schmidt or star formation law. The $\Sigma_{\rm SFR}$--$\Sigma_{\rm CO(1-0)}$ relation has a slope of $m=1.41\pm0.10$ when derived from the CO(1--0) data directly. For the CO(2--1) transition, the slope of the $\Sigma_{\rm SFR}$--$\Sigma_{\rm CO(2-1)}$ relation is shallower, $m=1.22\pm0.08$. If we compute the CO(1--0) relation using luminosities obtained by dividing our CO(2--1) data by a variable $r_{21}$, following the prescription of Equation~\ref{eq:r21prescription}, we obtain a slope of $m=1.35\pm0.09$, more consistent with CO(1--0). In a study of the the Kennicutt-Schidt law for resolved regions of 17 nearby galaxies, \citet{yajima+21} found that CO(2--1) gives slopes 10 to 20\% shallower than CO(1--0), in good agreement with our galaxy-integrated result. These results are also in rough agreement with the simulations of \citet{narayanan+11}, who predict a decrease in slope of $\sim0.1$ between the CO(1--0) and CO(2--1) relations. 

In Figure~\ref{fig:ms} we repeat this exercise for the correlations of ${\rm SFR} / L_{\rm CO}^\prime$ and $L_{\rm CO}^\prime / M_*$ with main sequence offset. We again observe differing slopes between the CO(2--1) and CO(1--0) relations and that these differences disappear when using Equation~\ref{eq:r21prescription} (Table~\ref{tab:galaxylaws}). 

The key point is that for perfectly matched samples of galaxies, the slope of scaling relations will differ depending on the choice of CO line unless variations of $r_{21}$ with galaxy properties are taken into account. These differences are subtle, but important. For example, the slopes of the Kennicutt-Schmidt law and main sequence offset--SFE correlation are taken as indicators of the relative importance of star formation efficiency versus gas abundance in determining the movement of galaxies above and bellow the main sequence \citep{saintonge+22} and in quenching star formation \citep{colombo+20}. The systematic differences between CO(2--1) and CO(1--0) are such that when gas mass is measured with CO(2--1) and a constant $r_{21}^c$, SFE will appear less important relative to gas abundance if the the same measurement was made with CO(1--0) directly.


\section{Conclusion}\label{sec:conclusion}

We present measurements of the CO(2--1)/CO(1--0) line ratio for 122 nearby galaxies using data from the AMISS and xCOLD~GASS projects. Our sample consists of a diverse sample of galaxies on, above, and below the star forming main sequence, spanning 2.5 orders of magnitude in both stellar mass and star formation rate. For the first time, this dataset allows us to measure correlations between the galaxy-integrated $r_{21}$ ratio and galaxy properties related to star formation activity with high confidence. 

Our key conclusions are as follows:
\begin{enumerate}[noitemsep,nolistsep]
    \item Galaxy-integrated $r_{21}$ ratios correlate strongly with a number of properties related to star formation. Galaxies with higher SFR, $\Sigma_{\rm SFR}$, SFE, main sequence offset, and sSFR show higher $r_{21}$ than galaxies with less intense star formation. Weaker correlations may also exist between $r_{21}$ and molecular gas mass or CO luminosity.
    \item The variations in galaxy-integrated $r_{21}$ can be described by the relation
    \begin{equation*}
        \log r_{21} =     
        \begin{dcases}
            0.12 \log {\rm SFR} - 0.19 & -1 \lesssim \log {\rm SFR} < 1.58 \\
            0.0 & 1.58 < \log {\rm SFR}
        \end{dcases}\,.
    \end{equation*}
    The intrinsic scatter around this relation is only $\sim$10\%.
    \item The large dynamic range of the AMISS data allows us to link the results of prior studies across four orders of magnitude in SFR. Literature $r_{21}$ values for nearby star forming galaxies, (ultra-)luminous infrared galaxies, and high redshift galaxies all lie along the $r_{21}$-SFR relation fitted to the AMISS data.
    \item $r_{21}$ has a limited dynamic range, and changes by only a factor of $\sim2$ over the full range of SFRs in the AMISS sample. This means the error caused by assuming a constant value of $r_{21}$ when using CO(2--1) to estimate gas mass is often small relative to the uncertainty in the $\alpha_{\rm CO}$ conversion factor.
    \item However, when comparing the gas content of galaxies with different star formation rates (or star formation surface densities, efficiencies, etc.), assuming a constant $r_{21}$ will introduce a systematic bias by overestimating the gas masses of galaxies undergoing more intense star formation and underestimating the gas masses of those forming fewer stars.
    \item For a matched sample of galaxies, the slopes scaling relations describing the relation between molecular gas and star formation in galaxies differ depending on whether CO(1--0) or CO(2--1) is used as a gas tracer. However, applying a variable $r_{21}$ based on our prescription for $r_{21}({\rm SFR})$ recovers consistent results for both CO(1--0) and CO(2--1).
\end{enumerate}

These results are highly robust to assumptions made in the data reduction and analysis (See Appendices~\ref{ap:crosscheck} and \ref{ap:12m}), and represent an investment of hundreds of hours of observation with the SMT and IRAM 30m. While further progress in unresolved studies of low-$J$ line ratios is possible by expanding our sample -- e.g. by including pushing to lower galaxy masses and metallicities -- our understanding of $r_{21}$ is now adequate to reliably substitute CO(2--1) for CO(1--0) when determining (galaxy-integrated) molecular gas masses in a wide range of contexts. Resolved $r_{21}$ measurements for a growing sample of very-nearby galaxies will further clarify the physical conditions driving variations in the CO line ratios, as well as reduce any remaining uncertainties associated with aperture corrections applied to our global CO measurements.

\acknowledgments
We wish to thank the Arizona Radio Observatory operators, engineering and management staff, without whom this research would not have been possible. We would also like to thank J. den Brok and J. Puschnig for sharing the compilation of literature $r_{21}$ values from \citet{denbrok+21}, and E. Mayer for helpful conversations during the development of this paper. RPK thanks J. Keenan for her tireless support. 

RPK was supported by the National Science Foundation through Graduate Research Fellowship grant DGE-1746060. DPM and RPK were supported by the National Science Foundation through CAREER grant AST-1653228.

This paper makes use of data collected by the the UArizona ARO Submillimeter Telescope and the UArizona ARO 12-meter Telescope, the IRAM 30m telescope, and the Sloan Digital Sky Survey. The UArizona ARO 12-meter Telescope on Kitt Peak  on Mt. Graham are operated by the Arizona Radio Observatory (ARO), Steward Observatory, University of Arizona. IRAM is supported by INSU/CNRS (France), MPG (Germany) and IGN (Spain). Funding for the SDSS and SDSS-II was provided by the Alfred P. Sloan Foundation, the Participating Institutions, the National Science Foundation, the U.S. Department of Energy, the National Aeronautics and Space Administration, the Japanese Monbukagakusho, the Max Planck Society, and the Higher Education Funding Council for England. The SDSS Web Site is \url{www.sdss.org}.

\appendix
\section{Cross-Checking of the $r_{21}$--SFR Scaling}\label{ap:crosscheck}

\begin{figure}
    \centering
    \includegraphics[width=.48\textwidth]{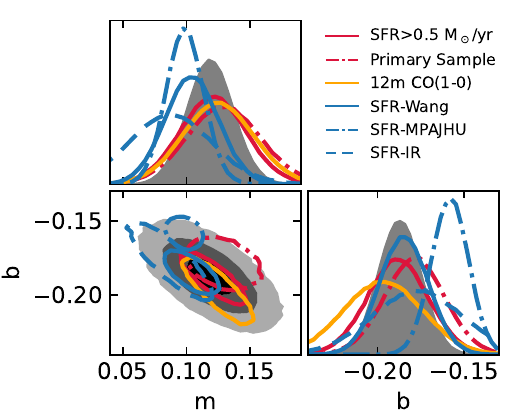}
    \caption{The posterior probability distribution for the parameters of a power law fit of $r_{21}$ versus SFR using different subsamples and datasets. Black/gray filled regions show fit results using our full $r_{21}$ sample and the SFRs provided by xCOLD~GASS. Colored lines show the results for subsamples (red), alternative measurements of $r_{21}$ (orange) and alternative SFR calibrations (blue). The upper left and lower right plots show the marginalized probability density for the slope $m$ and normalization $b$ terms of each fit. Contour lines in lower left plot show the 1$\sigma$ confidence regions for the joint distribution of $m$ and $b$. For the full sample, filled contours show 1, 2, and 3$\sigma$ confidence regions.}
    \label{fig:r21sfr_corner}
\end{figure}

\begin{deluxetable*}{l|c|c|cccc|c}
    \tablecaption{Correlation coefficients and regression parameters for alternative analyses of the $r_{21}$--SFR correlation.
    \label{tab:corr_alt}}
    \tablehead{
        \colhead{\textit{x}-variable} & \colhead{N} & \colhead{r} & \colhead{m} & \colhead{b} & \colhead{$\sigma$} & \colhead{$\rho_{mb}$} & \colhead{$m/\sigma_m$} \\
        \colhead{(1)} & \colhead{(2)} & \colhead{(3)} & \colhead{(4)} & \colhead{(5)} & \colhead{(6)} & \colhead{(7)} & \colhead{(8)}
    }
    \startdata
        \hline
        \multicolumn{8}{c}{Main Result (reproduced from Table~\ref{tab:corr})}\\ 
        \hline 
        ${\rm SFR}$ [${\rm M_\odot~yr^{-1}}$] & 120 & 0.487 & $0.119_{-0.018}^{+0.018}$ & $-0.187_{-0.012}^{+0.013}$ & $0.042_{-0.019}^{+0.016}$ & -0.62 & 6.5 \\
        \hline
        \multicolumn{8}{c}{Correlations and Fits using Alternative Samples/Data}\\ 
        \hline 
        ${\rm SFR}$ [${\rm M_\odot~yr^{-1}}$] (${\rm SFR}>0.5~{\rm M_\odot~yr^{-1}}$) & 99 & 0.409 & $0.121_{-0.026}^{+0.026}$ & $-0.188_{-0.018}^{+0.018}$ & $0.045_{-0.019}^{+0.016}$ & -0.81 & 4.6 \\
        ${\rm SFR}$ [${\rm M_\odot~yr^{-1}}$] (Primary Sample) & 52 & 0.473 & $0.129_{-0.029}^{+0.030}$ & $-0.179_{-0.018}^{+0.017}$ & $0.039_{-0.026}^{+0.028}$ & -0.31 & 4.4 \\
        ${\rm SFR}$ [${\rm M_\odot~yr^{-1}}$] (12m CO(1-0)) & 45 & 0.630 & $0.125_{-0.028}^{+0.028}$ & $-0.197_{-0.023}^{+0.023}$ & $0.023_{-0.015}^{+0.018}$ & -0.83 & 4.4 \\
        ${\rm SFR_{SED}}$ [${\rm M_\odot~yr^{-1}}$] & 117 & 0.398 & $0.104_{-0.021}^{+0.021}$ & $-0.185_{-0.014}^{+0.014}$ & $0.057_{-0.016}^{+0.015}$ & -0.65 & 4.9 \\
        ${\rm SFR_{MPAJHU}}$ [${\rm M_\odot~yr^{-1}}$] & 120 & 0.521 & $0.099_{-0.015}^{+0.015}$ & $-0.158_{-0.011}^{+0.011}$ & $0.021_{-0.015}^{+0.019}$ & -0.27 & 6.8 \\
        ${\rm SFR_{IR}}$ [${\rm M_\odot~yr^{-1}}$] & 54 & 0.304 & $0.087_{-0.033}^{+0.033}$ & $-0.177_{-0.027}^{+0.027}$ & $0.059_{-0.019}^{+0.019}$ & -0.85 & 2.7 \\
        ${\rm SFR}$ [${\rm M_\odot~yr^{-1}}$] (No Sys Errors) & 120 & 0.487 & $0.119_{-0.018}^{+0.019}$ & $-0.188_{-0.013}^{+0.013}$ & $0.068_{-0.010}^{+0.011}$ & -0.62 & 6.4    \enddata
    \tablecomments{Columns match Table~\ref{tab:corr}.}
\end{deluxetable*}

Expected values of $r_{21}$ span only a factor of 2-3, providing limited dynamic range over which to identify trends. Without very deep observations of both CO lines, the measurement noise for $r_{21}$ in an individual galaxy can be comparable in magnitude to the dynamic range of the trends between $r_{21}$ and other galaxy properties \citep{leroy+22}. The power of the AMISS sample lies in its large size, high dynamic range in relevant galaxy properties, and carefully calibrated data. However, even with high quality data, unidentified systematic errors can mask or confuse trends.

In this Appendix we report a number of checks to validate the correlation between SFR and $r_{21}$.  We have multiple independent SFR estimates and two semi-independent $r_{21}$ measurements, allowing us to explore multiple potential sources of bias in our results. These checks represent a test case for other trends presented in Section~\ref{ss:results-fit}, for which we do not have as many independent measurements, but which we expect to be subject to similar biases, if any exist. We find no evidence that sample selection or unaccounted for systematic errors bias our results, providing confidence that the trends we measure are astrophysical in origin.

Figure~\ref{fig:r21sfr_corner} shows the distribution of slope and normalization parameters of the fits performed for each test detailed below. We report the corresponding correlation coefficients and fit parameters in Table~\ref{tab:corr_alt}.

First, we refit the $r_{21}$--SFR correlation using only galaxies with SFR above our SFR completeness threshold of $0.5$~M$_\odot$~yr$^{-1}$ (in Figure~\ref{fig:r21sfr_corner} and Table~\ref{tab:corr_alt} this fit is labeled ``${\rm SFR}>0.5$''). This fit matches results for our full sample within $1\sigma$, indicating that selection effects for low SFR galaxies do not introduce a bias in fit results at ${\rm SFR}>0.5$~M$_\odot$~yr$^{-1}$. We cannot rule out the possibility that our sample is an incomplete representation of the behavior of $r_{21}$ for galaxies with SFRs between $0.08$ and $0.5$~M$_\odot$~yr$^{-1}$, however, the match between fits with and without these objects suggests that the low SFR galaxies lie on an extension of the trend found at higher SFRs. We also restricted our sample to only galaxies from our primary sample, which was selected to to uniformly cover the mass range $10^9<M_*<10^{11.5}$~M$_\odot$ with no consideration for SFR (label: ``Primary Sample''). Again we find near-perfect agreement between this fit and our main result, indicating that our inclusion of additional targets to probe the high-SFR end of the correlation has not biased our sample.

Next, we computed an alternative $r_{21}$ for 45 of our galaxies using a second CO(1--0) measurement from the ARO~12m telescope (label: ``12m CO(1--0)''; see Appendix~\ref{ap:12m}). The 30m and 12m CO(1--0) data were observed and reduced by completely independent groups using different facilities, and therefore comparison between these results provides a check on systematic uncertainties or biases in the reduction of our data. We find a nearly identical best fit.

Finally, we repeated our fits using alternative SFR calibrations based on SED fitting \citep[label: ``SFR-Wang'';][]{wang+11}, SDSS spectra \citep[label: ``SFR-MPAJHU'';][]{kauffmann+03,brinchmann+04}, or IR luminosity \citep[label: ``SFR-IR''; derived using the prescription of][]{kennicutt+12}. Each of these fits returns results consistent with our main result. The fit using IR luminosity-based SFRs allows a much wider range of slopes and the slope for this trend is nonzero at only $\sim3\sigma$ significance. However, IR luminosities are only available for 54 of our targets, and these span a fairly narrow range in SFR, we therefore expect that the $r_{21}$--${\rm SFR}_{\rm IR}$ fit would converge towards our other results with a larger sample and greater dynamic range. The remaining SFR tracers give similar fit results to the xCOLD~GASS SFRs, with identical slopes and slight offsets in normalization, which can most likely be attributed to systematic offsets between the various SFR calibrations.

\section{Cross-Checking the $r_{21}$--$L_{\rm CO}^\prime$ Scalings with Independent CO(1--0) Observations}\label{ap:12m}

A number of the $x$-variables considered in Section~\ref{ss:results-fit} are derived from CO luminosities. Without proper treatment, correlated errors in the $r_{21}$ and these $L_{\rm CO}^\prime$ derived quantities can alter the correlations recovered; for quantities with $L_{\rm CO}^\prime$ in the numerator, the slope will be artificially decreased, whereas for quantities with $L_{\rm CO}^\prime$ in the denominator, the slope will be increased. In the main text we have accounted for this in our fits by including a covariance term in the $x$- and $y$-errors. Here we report an alternative approach to characterizing this effect.

For 45 of our galaxies, we acquired additional CO(1--0) observations with the Arizona Radio Observatory's 12m ALMA prototype antenna. This allows us to use redundant CO(1--0) measurements to derive $x$- ($L_{\rm CO(1-0)}^\prime$, ${\rm SFR} / L_{\rm CO(1-0)}^\prime$, $\Sigma_{L_{\rm CO}^\prime}$, and $L_{\rm CO(1-0)}^\prime / M_*$) and $y$- ($r_{21}$) values with statistically independent errors.

Table~\ref{tab:corr_12m} reports correlation coefficients and power law fit parameters between $r_{21}$ and $L_{\rm CO(1-0)}^\prime$, ${\rm SFR} / L_{\rm CO(1-0)}^\prime$, $\Sigma_{L_{\rm CO}^\prime}$, and $L_{\rm CO(1-0)}^\prime / M_*$. For each $x$-variable we repeat our analysis in four ways:
\begin{enumerate}[noitemsep,nolistsep]
    \item using all galaxies in our sample and $x$- and $y$- variables derived from the IRAM~30m CO(1--0) data (the result from Section~\ref{ss:results-fit});
    \item using only galaxies re-observed with the ARO~12m, but still deriving both the $x$- and $y$- variables from the IRAM CO(1--0) data;
    \item using $x$-variables derived from the ARO~12m CO(1--0) data and $y$-variables derived from the IRAM~30m CO(1--0) data;
    \item using $x$-variables derived from the IRAM~30m CO(1--0) data and $y$-variables derived from the ARO~12m CO(1--0) data.
\end{enumerate}
Differences between Analysis 3/4 and Analysis 2 can be used to explore bias introduced by the correlated errors, while differences between Analysis 1 and Analysis 2 give a sense for how well our smaller (and less representative) 12m sample recovers trends seen in the full dataset.

For $L_{\rm CO(1-0)}^\prime$, $\Sigma_{L_{\rm CO}^\prime}$, and $L_{\rm CO(1-0)}^\prime / M_*$, the correlation coefficients and fitted slopes increase when using independent $x-$ and $y-$ variables. They decrease for ${\rm SFR} / L_{\rm CO(1-0)}^\prime$, although are consistent with our main result within the uncertainties.

These results suggest that the correlations between $r_{21}$ and molecular gas abundance (Section~\ref{sss:r21-gas}) may be just as strong as those between $r_{21}$ and SFR (Section~\ref{sss:r21-sfr}), but appear weaker due to correlated errors. On the other hand, the correlation between $r_{21}$ and star formation efficiency (${\rm SFR} / L_{\rm CO(1-0)}^\prime$), while real, may be overestimated somewhat because of correlated errors.

\begin{deluxetable*}{lccc|c|cccc|c}
    \tablecaption{Correlation coefficients and regression parameters for fits using different CO(1--0) data.
    \label{tab:corr_12m}}
    \tablehead{
        \colhead{\textit{x}-data} & \colhead{\textit{r}$_{21}$-data} & \colhead{Sample} & \colhead{N} & \colhead{r} & \colhead{m} & \colhead{b} & \colhead{$\sigma$} & \colhead{$\rho_{mb}$} & \colhead{$m/\sigma_m$} \\
        \colhead{(1)} & \colhead{(2)} & \colhead{(3)} & \colhead{(4)} & \colhead{(5)} & \colhead{(6)} & \colhead{(7)} & \colhead{(8)} & \colhead{(9)} & \colhead{(10)}
    }
    \startdata
        \hline
        \multicolumn{10}{c}{$r_{21}$--$L_{\rm CO}^\prime$ Correlations: $x$-variable $=L_{\rm CO}^\prime$ [$10^9~{\rm K~km~s^{-1}~pc^2}$]}\\ 
        \hline 
        IRAM 30m & IRAM 30m & Full Sample & 121 & 0.208 & $0.067_{-0.024}^{+0.024}$ & $-0.133_{-0.012}^{+0.012}$ & $0.067_{-0.015}^{+0.014}$ & 0.26 & 2.8 \\
        IRAM 30m & IRAM 30m & 12m Sample & 46 & 0.435 & $0.116_{-0.039}^{+0.040}$ & $-0.145_{-0.017}^{+0.017}$ & $0.064_{-0.021}^{+0.021}$ & -0.26 & 3.0 \\
        ARO 12m & IRAM 30m & 12m Sample & 46 & 0.542 & $0.143_{-0.036}^{+0.037}$ & $-0.147_{-0.016}^{+0.016}$ & $0.062_{-0.019}^{+0.019}$ & -0.25 & 3.9 \\
        IRAM 30m & ARO 12m & 12m Sample & 46 & 0.510 & $0.108_{-0.034}^{+0.034}$ & $-0.133_{-0.015}^{+0.014}$ & $0.040_{-0.019}^{+0.018}$ & -0.35 & 3.2 \\
        \hline
        \multicolumn{10}{c}{$r_{21}$--SFE Correlations: $x$-variable $={\rm SFR} / L_{\rm CO(1-0)}^\prime$ [$10^{-9}~{\rm M_\odot~yr^{-1}~(K~km~s^{-1}~pc^{2})^{-1}}$]}\\ 
        \hline 
        IRAM 30m & IRAM 30m & Full Sample & 120 & 0.474 & $0.173_{-0.029}^{+0.029}$ & $-0.228_{-0.019}^{+0.018}$ & $0.065_{-0.012}^{+0.012}$ & -0.83 & 6.0 \\
        IRAM 30m & IRAM 30m & 12m Sample & 45 & 0.556 & $0.201_{-0.046}^{+0.048}$ & $-0.229_{-0.029}^{+0.028}$ & $0.064_{-0.016}^{+0.017}$ & -0.85 & 4.4 \\
        ARO 12m & IRAM 30m & 12m Sample & 45 & 0.412 & $0.163_{-0.052}^{+0.053}$ & $-0.211_{-0.032}^{+0.031}$ & $0.066_{-0.019}^{+0.020}$ & -0.86 & 3.2 \\
        IRAM 30m & ARO 12m & 12m Sample & 45 & 0.380 & $0.119_{-0.046}^{+0.047}$ & $-0.174_{-0.028}^{+0.027}$ & $0.041_{-0.019}^{+0.018}$ & -0.85 & 2.6 \\
        \hline
        \multicolumn{10}{c}{$r_{21}$--$\Sigma_{L_{\rm CO}^\prime}$ Correlations: $x$-variable $=\Sigma_{L_{CO}^\prime}$ [${\rm K~km~s^{-1}}$]}\\ 
        \hline 
        IRAM 30m & IRAM 30m & Full Sample & 121 & 0.213 & $0.075_{-0.025}^{+0.025}$ & $-0.220_{-0.029}^{+0.028}$ & $0.064_{-0.016}^{+0.015}$ & -0.92 & 3.0 \\
        IRAM 30m & IRAM 30m & 12m Sample & 46 & 0.344 & $0.092_{-0.037}^{+0.038}$ & $-0.248_{-0.050}^{+0.049}$ & $0.078_{-0.018}^{+0.019}$ & -0.94 & 2.5 \\
        ARO 12m & IRAM 30m & 12m Sample & 46 & 0.460 & $0.117_{-0.034}^{+0.035}$ & $-0.278_{-0.046}^{+0.045}$ & $0.062_{-0.020}^{+0.020}$ & -0.94 & 3.5 \\
        IRAM 30m & ARO 12m & 12m Sample & 46 & 0.440 & $0.102_{-0.025}^{+0.025}$ & $-0.251_{-0.036}^{+0.035}$ & $0.020_{-0.014}^{+0.019}$ & -0.94 & 4.2 \\
        \hline
        \multicolumn{10}{c}{$r_{21}$--Gas Fraction Correlations: $x$-variable $= L_{\rm CO(1-0)}^\prime / M_*$ [${\rm K~km~s^{-1}~pc^{2}~M_\odot^{-1}}$]}\\ 
        \hline 
        IRAM 30m & IRAM 30m & Full Sample & 121 & 0.166 & $0.071_{-0.030}^{+0.030}$ & $-0.034_{-0.048}^{+0.047}$ & $0.068_{-0.015}^{+0.015}$ & 0.97 & 2.3 \\
        IRAM 30m & IRAM 30m & 12m Sample & 46 & 0.081 & $0.053_{-0.060}^{+0.059}$ & $-0.063_{-0.082}^{+0.080}$ & $0.078_{-0.021}^{+0.021}$ & 0.98 & 0.9 \\
        ARO 12m & IRAM 30m & 12m Sample & 46 & 0.273 & $0.125_{-0.058}^{+0.059}$ & $0.033_{-0.079}^{+0.080}$ & $0.073_{-0.020}^{+0.020}$ & 0.98 & 2.2 \\
        IRAM 30m & ARO 12m & 12m Sample & 46 & 0.380 & $0.132_{-0.038}^{+0.038}$ & $0.052_{-0.050}^{+0.051}$ & $0.025_{-0.017}^{+0.020}$ & 0.97 & 3.5 
        \enddata
    \tablecomments{Columns are (1) source of CO(1--0) data used to determine the $x$-variable; (2) source of CO(1--0) data used to determine $r_{21}$; (3) sample used; (4) number of galaxies considered; (5) Pearson correlation coefficient; (6-8) fit parameters and uncertainties; (9) correlation of the uncertainties in $m$ and $b$ ($\sigma_{mb}^2/\sigma_m \sigma_b$); (10) the ratio between the fitted slope and its uncertainty. \\
    Fits are of the form $\log r_{21} = m \log x + b + \mathcal{N}(\sigma)$ and are performed accounting for uncertainties in both $x$ and $y$.\\ 
    The values of $b$ are subject to an additional $0.03$~dex uncertainty due to uncertainty in the planet models used to set the flux scale for each telescope.}
\end{deluxetable*}

\bibliography{refs}

\end{document}